%% file: main.tex
\documentclass[
 reprint,
 eprint,
 groupedaddress,
 preprintnumbers,
 nofootinbib,
 amsmath,amssymb,
 aps,
 prd,
 floatfix,
]{revtex4-2}

\usepackage{graphicx}
\usepackage{dcolumn}
\usepackage{bm}
\usepackage{xcolor}
\definecolor{darkblue}{RGB}{46,48,147}
\usepackage{hyperref}
\hypersetup{colorlinks=true,
            linkcolor=darkblue,
            urlcolor=darkblue,
            citecolor=darkblue,
            breaklinks=true}
\usepackage{url}
\usepackage{booktabs}
\usepackage{topcapt}
\usepackage{chngpage}
\usepackage{xspace}
\usepackage{multirow}

\usepackage{array}
\usepackage{scalerel}
\usepackage{tikz}
\usetikzlibrary{svg.path}

\definecolor{orcidlogocol}{HTML}{A6CE39}
\tikzset{
  orcidlogo/.pic={
    \fill[orcidlogocol] svg{M256,128c0,70.7-57.3,128-128,128C57.3,256,0,198.7,0,128C0,57.3,57.3,0,128,0C198.7,0,256,57.3,256,128z};
    \fill[white] svg{M86.3,186.2H70.9V79.1h15.4v48.4V186.2z}
                 svg{M108.9,79.1h41.6c39.6,0,57,28.3,57,53.6c0,27.5-21.5,53.6-56.8,53.6h-41.8V79.1z M124.3,172.4h24.5c34.9,0,42.9-26.5,42.9-39.7c0-21.5-13.7-39.7-43.7-39.7h-23.7V172.4z}
                 svg{M88.7,56.8c0,5.5-4.5,10.1-10.1,10.1c-5.6,0-10.1-4.6-10.1-10.1c0-5.6,4.5-10.1,10.1-10.1C84.2,46.7,88.7,51.3,88.7,56.8z};
  }
}

\newcommand\orcidicon[1]{\href{https://orcid.org/#1}{\mbox{\scalerel*{
\begin{tikzpicture}[yscale=-1,transform shape]
\pic{orcidlogo};
\end{tikzpicture}
}{0}}}}

\newcommand{\PreserveBackslash}[1]{\let\temp=\\#1\let\\=\temp}
\newcolumntype{C}[1]{>{\PreserveBackslash\centering}p{#1}}

\renewcommand{\vec}[1]{\mathbf{#1}}

\newcommand{\pt}{\ensuremath{p_{\mathrm{T}}}\xspace}

\newcommand{\TeV}{\ensuremath{\,\text{Te\hspace{-.08em}V}}\xspace}

\newcommand{\real}[1]{\ensuremath{#1_\mathrm{real}}\xspace}
\newcommand{\gen}[1]{\ensuremath{#1_\mathrm{gen}}\xspace}
\newcommand{\fgdinf}{\ensuremath{\mathrm{FGD}_\infty}\xspace}

\newcommand{\jetnet} {{\textsc{JetNet}}\xspace}

\newcommand{\etarel}{\ensuremath{\eta^\mathrm{rel}}\xspace}
\newcommand{\phirel}{\ensuremath{\phi^\mathrm{rel}}\xspace}
\newcommand{\ptrel}{\ensuremath{\pt^\mathrm{rel}}\xspace}

\newcommand{\wassm}{\ensuremath{W_1^\mathrm M}\xspace}
\newcommand{\wassppt}{\ensuremath{W^{\ptrel}_{1p}}\xspace}

\begin{document}

\preprint{FERMILAB-PUB-22-872-CMS-PPD}

\title{Evaluating generative models in high energy physics}

\author{Raghav Kansal\,\orcidicon{0000-0003-2445-1060}}
 \altaffiliation[Also at ]{Fermilab}
 \email{rkansal@ucsd.edu}
\author{Anni Li\,\orcidicon{0000-0002-7989-2894}}
\author{Javier Duarte\,\orcidicon{0000-0002-5076-7096}}
\affiliation{
 University of California San Diego, La Jolla, CA 92093, USA
}

\author{Nadezda Chernyavskaya\,\orcidicon{0000-0002-2264-2229}, Maurizio Pierini\,\orcidicon{0000-0003-1939-4268}}
\affiliation{
 European Center for Nuclear Research (CERN), 1211 Geneva 23, Switzerland
}

\author{Breno Orzari\,\orcidicon{0000-0003-4232-4743}, Thiago Tomei\,\orcidicon{0000-0002-1809-5226}}
\affiliation{
 Universidade Estadual Paulista, S\~{a}o Paulo/SP, CEP 01049-010, Brazil
 \vspace{\baselineskip}
}

\received{18 November 2022}
\accepted{8 March 2023}
\published{21 April 2023}

\begin{abstract}
There has been a recent explosion in research into machine-learning-based generative modeling to tackle computational challenges for simulations in high energy physics (HEP). 
In order to use such alternative simulators in practice, we need well-defined metrics to compare different generative models and evaluate their discrepancy from the true distributions.
We present the first systematic review and investigation into evaluation metrics and their sensitivity to failure modes of generative models, using the framework of two-sample goodness-of-fit testing, and their relevance and viability for HEP.
Inspired by previous work in both physics and computer vision, we propose two new metrics, the Fr\'echet and kernel physics distances (FPD and KPD, respectively), and perform a variety of experiments measuring their performance on simple Gaussian-distributed, and simulated high energy jet datasets.
We find FPD, in particular, to be the most sensitive metric to all alternative jet distributions tested and recommend its adoption, along with the KPD and Wasserstein distances between individual feature distributions, for evaluating generative models in HEP.
We finally demonstrate the efficacy of these proposed metrics in evaluating and comparing a novel attention-based generative adversarial particle transformer to the state-of-the-art message-passing generative adversarial network jet simulation model.
The code for our proposed metrics is provided in the open source \jetnet Python library.\\\\
DOI: \href{https://doi.org/10.1103/PhysRevD.107.076017}{10.1103/PhysRevD.107.076017}
\end{abstract}

\maketitle

%\tableofcontents

\section{\label{sec:intro} Introduction}

In high energy physics (HEP), accurate simulations are critical for precision measurements and searches such as those performed at the CERN Large Hadron Collider (LHC). 
These are traditionally performed using Monte Carlo (MC) event generators, detailed modeling of particles' propagation and interaction through detectors (typically with the GEANT4~\cite{geant4} package), and reconstruction algorithms to unfold detector measurements back to particles and high-level objects such as jets. 
While these methods have been highly successful for the physics goals of the LHC, scaling up to the simulation challenges of the upcoming high-luminosity phase of the LHC~\cite{HL-LHC} necessitates significant advancements in speed and resource requirements~\cite{CMS:2815292,ATLAS_HLLHC,HEPSoftwareFoundation:2020daq}, while maintaining the quality of current simulations. 

To tackle this problem, a plethora of techniques for fast simulation of calorimeter showers and jets have been developed and explored in the past few years, particularly using generative modeling techniques in machine learning (ML)~\cite{Paganini:2017dwg,deOliveira:2017pjk,Paganini:2017hrr,Butter:2019cae,Belayneh:2019vyx,DiSipio:2019imz,Hashemi:2019fkn,Carrazza:2019cnt,Carminati:2020kym,Gao:2020zvv,Buhmann:2021caf,kansal_mpgan,krause_caloflow,Bieringer:2022cbs,Touranakou:2022qrp,Mikuni:2022xry,ATLAS:2022jhk}. 
Reviews of these approaches can be found in Refs.~\cite{Butter:2020tvl,Alanazi:2021grv}.
For an experimental collaboration to apply one of these techniques in real data analyses, however, they require methods to objectively compare the performance of different simulation techniques and extensively validate the produced simulations. 
This calls for the study and adoption of standard quantitative evaluation metrics for generative modeling in HEP. 

Recently, several metrics have been proposed to address this challenge.
However, to our knowledge, there has been no systematic investigation of their sensitivity to expected failure modes of generative models, and their relevance to validation and feasibility for broad adoption in HEP. 
To this end, we study the performance of several proposed metrics from HEP and computer vision.
Inspired by both domains, we develop two novel metrics we call the Fr\'echet and kernel physics distances (FPD and KPD, respectively) and find them to collectively have excellent sensitivity to all tested data mismodeling, as well as to satisfy practical requirements for evaluation and comparison of generative models in HEP.
We conclude our experiments by recommending the adoption of FPD and KPD, along with quantifying differences in individual feature distributions using the Wasserstein 1-distance, and demonstrate their use in evaluating a novel attention-based generative model we call the generative adversarial particle transformer (GAPT).
We provide implementations for the new metrics in the \jetnet library~\cite{kansal_jetnet_library}.

This paper is structured as follows. 
In Sec.~\ref{sec:metrics} we define our criteria for evaluation metrics in HEP and review existing metrics. 
We present results on the performance of these metrics on Gaussian-distributed synthetic toy data and simulated high energy jets in Secs.~\ref{sec:toydata} and~\ref{sec:jetdata} respectively. 
Based on these experiments, we provide our recommendations and concretely illustrate their application by evaluating and comparing GAPT to the current state-of-the-art (SOTA) MPGAN~\cite{kansal_mpgan} model in Sec.~\ref{sec:gapt}.
Finally, we conclude in Sec.~\ref{sec:conclusion}.

\section{\label{sec:metrics} Evaluation metrics for generative models}

In evaluating generative models, we aim to quantify the difference between the real and generated data distributions $\real{p}(\vec{x})$ and $\gen{p}(\vec{x})$ respectively, where data samples $\vec{x} \in \mathbb{R}^d$ are typically high dimensional.
Lacking tractable analytic distributions in general, this can be viewed as a form of two-sample goodness-of-fit (GOF) testing of the hypothesis $\real{p}(\vec{x}) = \gen{p}(\vec{x})$ using real and generated samples, $\{\vec{\real{x}}\}$ and $\{\vec{\gen{x}}\}$, drawn from their respective distributions.
As illustrated in Ref.~\cite{cousins_gof}, in general, there is no ``best'' GOF test with power against all alternative hypotheses. 
Instead, we aim for a set of tests that collectively have power against the relevant alternatives we expect, and are practically most appropriate.
We first outline the criteria we require of our evaluation metrics in Sec.~\ref{sec:criteria}, then review and discuss the suitability of possible metrics in Sec.~\ref{sec:emetrics}, and end in Sec.~\ref{sec:feature_selection} with a discussion on the features to use in comparing such high-dimensional distributions, thereby motivating FPD and KPD.

\subsection{\label{sec:criteria} Criteria for evaluation metrics in HEP}

Typical failure modes in ML generative models such as normalizing flows and autoregressive models include a lack of sharpness and smearing of low-level features, while generative adversarial networks (GANs) often suffer from ``mode collapse'', where they fail to capture the entire real distribution, only generating samples similar to a particular subset. 
Therefore, with regard to the performance of generative models, we require first and foremost that the tests be sensitive to both the quality and the diversity of the generated samples.
It is critical that these tests are multivariate as well, particularly when measuring the performance of conditional models, which learn conditional distributions given input features such as those of the incoming particle into a calorimeter or originating parton of a jet, and which will be necessary for applications to LHC simulations~\cite{Butter:2022rso}.
Multivariate tests are required in order to capture the correlations between different features, including those on which such a model is conditioned. 
% note about multivariate g.o.f testing?
Finally, it is desirable for the test's results to be interpretable to ensure trust in the simulations. 

To facilitate a fair, objective comparison between generative models, we also require the tests to be reproducible---i.e., repeating the test on a fixed set of samples should produce the same result---and standardizable across different datasets, such that the same test can be used for multiple classes and data structures (e.g., both images and point clouds for calorimeter showers or jets).
It is also desirable for the test to be reasonably efficient in terms of speed and computational resources, to minimize the burden on researchers evaluating their models.

\subsection{\label{sec:emetrics} Evaluation metrics}

Having outlined criteria for our metrics, we now discuss possible metrics and their merits and limitations.
The traditional method for evaluating simulations in HEP is to compare physical feature distributions using one-dimensional (1D) binned projections. 
This allows valuable, interpretable insight into the physics performance of these simulators.
% , and can be quantified by measures such as the binned chi-squared ($\chi^2$) test.
% or the unbinned Wasserstein $p$-distance ($W_p$)~\cite{wasserstein_original,villani_ot}.
However, it is intractable to extend this binned approach to multiple distributions simultaneously, as it falls victim to the curse of dimensionality---the number of bins and samples required to retain a reasonable granularity in our estimation of the multidimensional distribution grows exponentially with the number of dimensions.
Therefore, while valuable, this method is restricted to evaluating single features, losing sensitivity to correlations and conditional distributions.

\subsubsection{\label{sec:ipmsfdivs} Integral probability metrics and \texorpdfstring{$f$}{f}-divergences}

To extend to multivariate distributions, we first review measures of differences between probability distributions.
The two prevalent, almost mutually exclusive,\footnote{The total variation distance is the only nontrivial discrepancy measure that is both an IPM and an $f$-divergence~\cite[Appendix A]{sriperumbudur_empirical}; however, to our knowledge, a consistent finite-sample estimator for it does not exist (see, for example, Ref.~\cite[Sec. 5]{sriperumbudur_empirical}).} classes of discrepancy measures are integral probability metrics (IPMs)~\cite{muller_ipms} and $f$-divergences.
An IPM $D_\mathcal{F}$, defined as
\
\begin{equation}\label{eqn:ipm}
    D_\mathcal{F}(\real p, \gen p) = \sup _{f \in \mathcal F} |\mathbb E_{\vec x \sim p_\mathrm{real}} f(\vec{x}) - \mathbb E_{\vec y \sim p_\mathrm{gen}} f(\vec y)|,
\end{equation}
measures the difference in two distributions, \real{p} and \gen{p} in Eq.~(\ref{eqn:ipm}), by using a ``witness'' function $f$, out of a class of measurable, real-valued functions $\mathcal{F}$, which maximizes the absolute difference in its expected value over the two distributions.
The choice of $\mathcal{F}$ defines different types of IPMs.
The famous Wasserstein 1-distance ($W_1$)~\cite{wasserstein_original,villani_ot}, for example, is an IPM for which $\mathcal F$ in Eq.~(\ref{eqn:ipm}) is the set of all $K$-Lipschitz functions (where $K$ is any positive constant).
Maximum mean discrepancy (MMD)~\cite{gretton_mmd} is another popular example, where $\mathcal F$ is the unit ball in a reproducing kernel Hilbert space (RKHS).

$f$-divergences, on the other hand, are defined as
\
\begin{equation}\label{eqn:fdiv}
    D_f(\gen p, \real p) = \int \real p (\vec x) f \bigg( \cfrac{\real p (\vec x)}{\gen p (\vec x)} \bigg ) d\vec x.
\end{equation}
They calculate the average of the pointwise differences between the two distributions, \real{p} and \gen{p} in Eq.~(\ref{eqn:ipm}), transformed by a ``generating function'' $f$, weighted by \real{p}.
Like IPMs, different $f$-divergences are defined by the choice of generating function.
Famous examples include the Kullback-Leibler (KL)~\cite{kl} and Jenson-Shannon (JS)~\cite{js_1,js_2} divergences, which are widely used in information theory to capture the expected information loss when modeling \real{p} by \gen{p} (or vice versa), as well as the Pearson $\chi^2$~\cite{pearson} divergence and related metrics~\cite{Baker:1983tu,generalization_gof,parametric}, which are ubiquitous in HEP as GOF tests.

% IPMs include popular measures such as the Wasserstein 1-distance ($W_1$) ~\cite{wasserstein_original,villani_ot}, maximum mean discrepancy (MMD)~\cite{gretton_mmd}, and energy distance~\cite{rizzo_energydist}, while measures such as the Kullback--Leibler (KL)~\cite{kl}, Jenson--Shannon (JS)~\cite{js_1,js_2}, and Pearson $\chi^2$~\cite{pearson} divergences are all forms of $f$-divergences. 
% These two, almost mutually exclusive%
% , 
% classes each have unique, interesting properties.

Overall, $f$-divergences can be powerful measures of discrepancies, with convenient information-theoretic interpretations and the advantage of coordinate invariance.
However, unlike IPMs, they do not generally take into account the metric space of distributions, because of which we argue that IPMs are more useful for evaluating generative models and their respective learned distributions.
An illustrative example of this is provided in Appendix~\ref{app:metricspace}.
IPMs can thereby be powerful metrics with which to compare different models, with measures such as $W_1$ and MMD able to metrize the weak convergence of probability measures~\cite{villani_ot, simongabriel_mmdmetrizing}.

% On the other hand, in general, $f$-divergences do not take the metric space into account, and instead measure pointwise differences in probability mass~\cite{renyi_measures}.
% This allows for coordinate invariance and convenient information-theoretic interpretations.
% $f$-divergences can thus be powerful measures of discrepancies, with the $\chi^2$ GOF test and related variants~\cite{Baker:1983tu,generalization_gof,parametric} ubiquitous in HEP.
% However, because they do not generally take into account the metric space of distributions, we argue that IPMs are more useful for \textit{comparing} generative models and their respective learnt distributions.

Additionally, on the practical side, finite-sample estimation of $f$-divergences such as the KL and the Pearson $\chi^2$ divergences is intractable in high dimensions, generally requiring partitioning in feature space, which suffers from the curse of dimensionality as described above.
References~\cite{sriperumbudur_empirical, sriperumbudur_ipms} demonstrate more rigorously the efficacy of finite-sample estimation of IPMs, in comparison to the difficulty of estimating $f$-divergences.

\subsubsection{\label{sec:ipms} IPMs as evaluation metrics}

Having argued in their favor, we discuss specific IPMs and related measures, and their viability as evaluation metrics. 
The most famous is the Wasserstein distance~\cite{wasserstein_original,villani_ot}, as defined above.
It is closely related to the problem of optimal transport~\cite{villani_ot}: finding the minimum ``cost'' to transport the mass of one distribution to another, when the cost associated with the transport between two points is the Euclidean distance between them.
This metric is sensitive to both the quality and diversity of generated distributions; however, its finite-sample estimator is the optimum of a linear program---an optimisation problem with linear constraints and objective~\cite{vanderbei2013linear}, which, while tractable in 1D, is biased with very poor convergence in high dimensions~\cite{ramdas_wasserstein}.
We demonstrate these characteristics empirically in Secs.~\ref{sec:toydata} and~\ref{sec:jetdata}.

A related \textit{pseudometric}\footnote{This is a pseudometric because distinct distributions can have a distance of 0 if they have the same means and covariances.} is the Fr\'echet, or $W_2$, distance between Gaussian distributions fitted to the features of interest, which we generically call the Fr\'echet Gaussian distance (FGD).
A form of this known as the Fr\'echet \textsc{Inception} distance (FID)~\cite{TTUR}, using the activations of the \textsc{Inception} v3 convolutional neural network model~\cite{inception_v3} on samples of real and generated images as its features, is currently the standard metric for evaluation in computer vision.
The FID has been shown to be sensitive to both quality and mode collapse in generative models and is extremely efficient to compute; however, it has the drawback of assuming Gaussian distributions for its features.
While finite-sample estimates of the FGD are biased~\cite{binkowski_demystifying}, Ref.~\cite{chong_unbiasedfid} introduces an effectively unbiased estimator \fgdinf, obtained by extrapolating from multiple finite-sample estimates to the infinite-sample value. 

The final IPM we discuss is the MMD~\cite{mmd}, for which $\mathcal F$ is the unit ball in an RKHS for a chosen kernel $k(x, y)$.
Intuitively, it is the distance between the mean embeddings of the two distributions in the RKHS, and it has been demonstrated to be a powerful two-sample test~\cite{gretton_mmd, liu_deepkernels}.
However, generally, high sensitivity requires tuning the kernel based on the two sets of samples.
For example, the traditional choice is a radial basis function kernel, where kernel bandwidth is typically chosen based on the statistics of the two samples~\cite{gretton_mmd}.
While such a kernel has the advantage of being characteristic---i.e., it produces an injective embedding~\cite{sriperumbudur_rkhs}---to maintain a standard and reproducible metric, we experiment instead with fixed polynomial kernels of different orders.
These kernels allow access to high order moments of the distributions and have been proposed in computer vision as an alternative to FID, termed kernel \textsc{Inception} distance (KID)~\cite{binkowski_demystifying}.
MMD has unbiased estimators~\cite{gretton_mmd}, which have shown to converge quickly even in high dimensions~\cite{binkowski_demystifying}.

\subsubsection{Manifold estimation}

Another form of evaluation metrics recently popularized in computer vision involves estimating the underlying manifold of the real and generated samples.
While computationally challenging, such metrics can be intuitive and allow us to disentangle the aspects of quality and diversity of the generated samples, which can be valuable in diagnosing individual failure modes of generative models.
The most popular metrics are ``precision'' and ``recall'' as defined in Ref.~\cite{kynkaanniemi_pr}.
For these, manifolds are first estimated as the union of spheres centered on each sample with radii equal to the distance to the $k$th-nearest neighbor.
Precision is defined as the number of generated points which lie within the real manifold, and recall as the number of real points within the generated manifold.
Alternatives, named diversity and coverage, are proposed in Ref.~\cite{naeem_dc} with a similar approach, but which use only the real manifold, and take into account the density of the spheres rather than just their union.
We study the efficacy of both pairs of metrics for our problem in Secs.~\ref{sec:toydata} and~\ref{sec:jetdata}.

\subsubsection{Classifier-based metrics}

Finally, an alternative class of GOF tests proposed in Refs.~\cite{friedman_gof, lopez_paz_c2st, liu_deepkernels}, and most relevantly in Ref.~\cite{krause_caloflow} and the fast calorimeter simulation challenge~\cite{calochallenge} to evaluate simulated calorimeter showers, are based on binary classifiers trained between real and generated data.
These tests have been posited to have sensitivity to both quality and diversity; however, they have significant practical and conceptual drawbacks in terms of understanding and comparing generative models. 

First, deep neural networks (DNNs) are widely considered uninterpretable black boxes~\cite{black_box}, hence it is difficult to discern which features of the generated data the network is identifying as discrepant or compatible.
Second, the performance of DNNs is highly dependent on both the architecture and dataset, and it is unclear how to specify a standard architecture sensitive to all possible discrepancies for all datasets.
Furthermore, training of DNNs is typically stochastic, minimizing a complex loss function with several potential local minima, and slow; hence it is sensitive to initial states and hyperparameters irrelevant to the problem, difficult to reproduce, and not efficient.

In terms of GOF testing, evaluating the performance of an individual generative model requires a more careful understanding of the null distribution of the test statistic than is proposed in Refs.~\cite{krause_caloflow, calochallenge}, such as by using a permutation test as suggested in Refs.~\cite{friedman_gof, liu_deepkernels} or retraining the model numerous times between samples from the true distribution as proposed recently in Refs.~\cite{dagnolo_nplm, dagnolo_lmnp} with applications to HEP searches.
However, even if such a test was performed for each model, which would itself be practically burdensome, it would remain difficult to fairly compare models, as, since different classifiers are trained for each model, this means comparing values of entirely different test statistics.\footnote{In the case of Refs.~\cite{dagnolo_nplm, dagnolo_lmnp} the test statistic remains the same, but estimating the null distribution is even more practically challenging, as it involves multiple trainings of the classifier.}
% As a corollary to these issues, if different binary classifiers are trained for each model, it begs the question of whether classifier summary metrics, e.g. the area under the curve (AUC) and accuracy, obtained for different models can be fairly compared at all.
% Finally, there is no straightforward way to extend this method to conditional evaluation, apart from binning in the conditioning variables, which faces the curse of dimensionality and leads to an even more computationally challenging metric.
Despite these drawbacks, we perform the classifier-based test from Refs.~\cite{krause_caloflow, calochallenge} in Sec.~\ref{sec:jetdata} and find that, perhaps surprisingly, it is insensitive to a large class of failures typical of ML generative models.

\subsection{\label{sec:feature_selection} Feature selection}

We end this section by discussing which features to select for evaluation. 
Generally, for data such as calorimeter showers and jets, individual samples $\vec x \in \mathbb R^d$ are extremely high dimensional, with showers and jets containing up to $\mathcal{O}(1000)$s of hits and particles respectively, each with its own set of features. 
Apart from the practical challenges of comparing distributions in this $d$-dimensional case, often this full set of low-level features is not the most relevant for our downstream use case.

This is an issue in computer vision as well, where images are similarly high dimensional, and comparing directly the low-level, high-dimensional feature space of pixels is not practical or meaningful.
Instead, the current solution is to derive salient, high-level features from the penultimate layer of a pretrained SOTA classifier.

This approach is necessary for images, for which it is difficult to define such meaningful numerical features by hand.
It can also be used in HEP as in Ref.~\cite{kansal_mpgan}, which proposed the Fr\'echet ParticleNet distance (FPND), using the ParticleNet~\cite{qu_particlenet} jet classifier to derive its features.
However, one key insight and study of this work is that this may be unnecessary for HEP applications, as we have already developed a variety of meaningful, hand-engineered features such as jet observables~\cite{marzani_jets, larkoski_jets, komiske_efps} and shower-shape variables~\cite{baffioni_electronrecocms, atlas_photonrecoatlas}.
Such variables may lead to a more efficient, more easily standardized, and interpretable test.
We experiment with both types of features in Sec.~\ref{sec:jetdata}. 

\section{\label{sec:toydata} Experiments on Gaussian-Distributed Data}

As a first test and filtering of the many metrics discussed, we evaluate each metric's performance on simple 2D (mixture of) Gaussian-distributed datasets. 
We describe the specific metrics tested in Sec.~\ref{sec:toydata_metrics}, the distributions we evaluate in Sec.~\ref{sec:toydata_distributions}, and experimental results in Sec.~\ref{sec:toydata_results}. 

\subsection{{\label{sec:toydata_metrics} Metrics}}

We test several metrics discussed in Sec.~\ref{sec:metrics}, with implementation details provided below. 
Values are measured for different numbers of test samples, using the mean of five measurements each and their standard deviation as the error, for all metrics but \fgdinf and MMD.
The sample size was increased until the metric was observed to have converged, or, as in the case of the Wasserstein distance and diversity and coverage, until it proved too computationally expensive.
Timing measurements for each metric can be found in Appendix~\ref{app:details}.

\begin{enumerate}
    \item \textbf{Wasserstein distance} is estimated by solving the linear program described in, for example, Ref.~\cite{bertsimas_linearopt}, using the Python optimal transport library~\cite{flamary_pot}.
    \item $\mathbf{\textbf{FGD}}_\infty$ is calculated by measuring FGD for 10 batch sizes, between a minimum batch size of 20,000 and varying maximum batch size. 
    A linear fit is performed of the FGD as a function of the reciprocal of the batch size, and \fgdinf is defined to be the $y$ intercept---it, thus, corresponds to the infinite batch size limit.
    The error is taken to be the standard error of the intercept.
    % The FGD at each of the 10 batch sizes is measured 20 times and the average is used for the linear fit.
    This closely follows the recommendation of Ref.~\cite{chong_unbiasedfid}, except empirically we find it necessary to increase the minimum batch size from 5,000 to 20,000 and to use the average of 20 measurements at each batch size in the linear fit, in order to obtain \fgdinf intervals with $>$68\% coverage of the true value.\footnote{The tests of coverage are performed on the jet distributions described in Sec.~\ref{sec:jetdata}, with the true FGD estimated as the FGD between batch sizes of 150,000, similar to Ref.~\cite{chong_unbiasedfid}.}
    \item \textbf{MMD} is calculated using the unbiased quadratic time estimator defined in Ref.~\cite{gretton_mmd}. 
    We test 3rd (as in KID) and 4th order polynomial kernels.
    We find MMD measurements to be extremely sensitive to outlier sets of samples, hence we use the median of 10 measurements each per sample size as our estimates, and half the difference between the 16th and 84th percentile as the error.
    We find empirically that this interval has 74\% coverage of the true value when testing on the true distribution.
    \item \textbf{Precision and recall}~\cite{kynkaanniemi_pr} and
    \item \textbf{Diversity and coverage}~\cite{naeem_dc} are both calculated using the recommendations of their respective authors, apart from the maximum batch size, which we vary.
\end{enumerate}

\subsection{{\label{sec:toydata_distributions} Distributions}}

\begin{figure*}[htbp]
    \includegraphics[width=\textwidth]{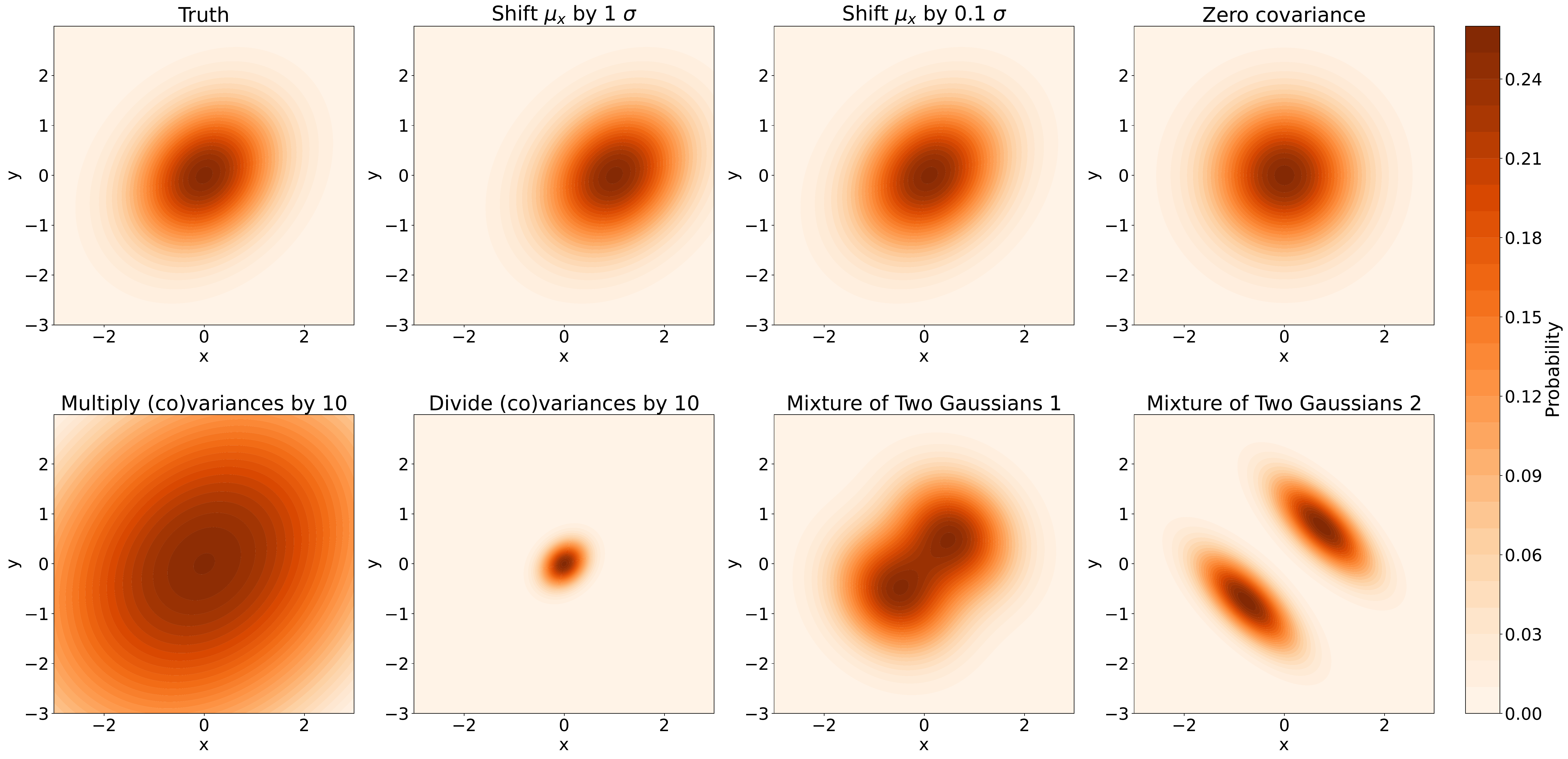}
    \caption{Samples of (mixtures of) Gaussian distributions used for testing evaluation metrics.}
    \label{fig:toydists}
\end{figure*}

\begingroup  % increase spacing for line with matrix in it.
\setlength{\lineskip}{5pt} 

We use a 2D Gaussian with 0 means and covariance matrix $\Sigma = \left(\begin{smallmatrix}
    1.00 & 0.25 \\ 0.25 & 1.00
\end{smallmatrix}\right)$ as the true distribution.
We test the sensitivity of the above metrics to the following distortions, shown in Fig.~\ref{fig:toydists}:

\endgroup

\begin{enumerate}
    \item a large shift in $x$ (1 standard deviation $\sigma$);
    \item a small shift in $x$ ($0.1\,\sigma$);
    \item removing the covariance between the parameters---this tests the sensitivity of each metric to correlations;
    \item multiplying the (co)variances by 10---tests sensitivity to quality;
    \item dividing (co)variances by 10---tests sensitivity to diversity; and, finally,
    \item[6 \& 7.] two mixtures of two Gaussian distributions with the same combined means, variances, and covariances as the truth---this tests sensitivity to the shape of the distribution.
\end{enumerate}

\subsection{{\label{sec:toydata_results} Results}}

\begin{figure*}[ht]
    \includegraphics[width=\textwidth]{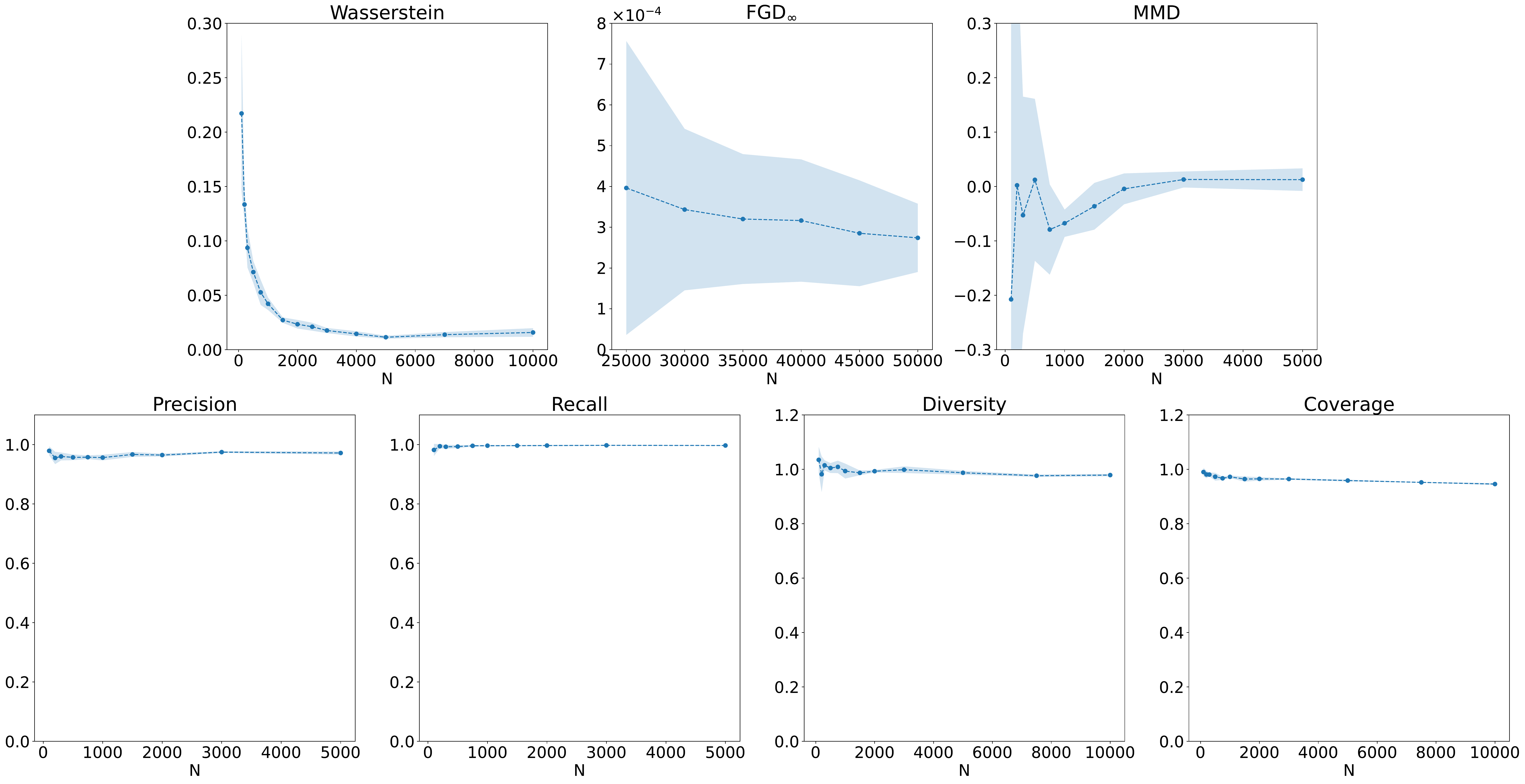}
    \caption{Scores of each metric on samples from the true distribution for varying sample sizes.}
    \label{fig:toy_truth_scores}
\end{figure*}

\subsubsection{Bias}

We first discuss the performance of each metric in distinguishing between two sets of samples from the truth distribution in Fig.~\ref{fig:toy_truth_scores}, effectively estimating the null distributions of each test statistic.
A fourth-order polynomial kernel for MMD is shown as it proved most sensitive.
We see that indeed \fgdinf and MMD are effectively unbiased, while the values of others depend on the sample size. 
This is a significant drawback; even if the same number of samples is specified for each metric to mitigate the effect of the bias, as discussed in Ref.~\cite{chong_unbiasedfid}, in general there is no guarantee that the level of bias \textit{for a given sample size} is the same across different distributions.
One possible solution is to use a sufficiently large number of samples to ensure convergence within a certain percentage of the true value. 
However, from a practical standpoint, the Wasserstein distance quickly becomes computationally intractable beyond $\mathcal{O}(1000)$ samples, before which, as we see in Fig.~\ref{fig:toy_truth_scores}, it does not converge even for a two-dimensional distribution.
Similarly, diversity and coverage require a large number of samples for convergence, which is impractical given their $\mathcal{O}(n^2)$ scaling, while precision and recall suffer from the same scaling but converge faster.

\subsubsection{Sensitivity}

\begin{table*}[t]
\topcaption{Values, significances, and errors of metrics, as defined in Sec.~\ref{sec:toydata_metrics}, for each (mixture of) Gaussian distribution(s), for the largest sample size tested. \label{tab:toy_results} 
The most significant scores per distribution are in bold.
}
\begin{adjustwidth}{-1in}{-1in}% adjust the L and R margins by 1 inch
\centering\resizebox{\textwidth}{!}{
    \begin{tabular}{l|C{2cm}C{2cm}C{2cm}C{2cm}C{2cm}C{2cm}C{2cm}C{2cm}}
    \toprule
    \input{tables/1_toy_measurements}\\
    \bottomrule
    \end{tabular}
}
\end{adjustwidth}
\end{table*}

Table~\ref{tab:toy_results} lists the means and errors of each metric per dataset for the largest sample size tested for each.
A similar plot to Fig.~\ref{fig:toy_truth_scores} for each alternative distribution can be found in Appendix~\ref{app:details}.
A significance is also calculated for each score by assuming a Gaussian null (truth) distribution,\footnote{We note that this is not necessarily the case, particularly for the Wasserstein distance, which has a biased estimator. 
However, this is not a significant limitation, because, as can be seen in Table~\ref{tab:toy_results}, there is rarely a significant overlap between the null and alternative distributions which would require an understanding of the shape of the former.}
and the most significant scores per alternative distribution are highlighted in bold.
% This is conceptually equivalent to assuming a, and highlighting the test statistic producing a central value with the highest $p$-value per alternative distribution.
We can infer several properties of each metric from these measurements.
% There are a lot of interesting conclusions we can draw from these tests on simple toy distributions.

Focusing first on the holistic metrics (Wasserstein, \fgdinf, and MMD), we find that each converges to ${\approx}0$ on the truth distribution, indicating their estimators are consistent.
We can evaluate the sensitivity to each alternative distribution by considering the difference in scores versus the truth scores.
With the notable exception of \fgdinf on the mixtures of two Gaussian distributions, we observe that all three metrics find the alternatives discrepant from the truth score with a significance of $>$2 (equivalent to a $p$-value of $<$0.05 of the test statistic on the alternative distributions).

As expected, despite the clear difference in the shapes of the mixtures compared to the truth, since \fgdinf has access to up to only the second-order moments of the distributions, it is not sensitive to such shape distortions.
We also note that a fourth-order polynomial kernel, as opposed to the third-order kernel proposed for KID, is required for MMD to be sensitive to the mixtures of Gaussian distributions, as shown in Appendix~\ref{app:details}.
\fgdinf is, however, generally the most sensitive to other alternative distributions.

Finally, we note that precision and recall are clearly sensitive to the two distributions designed to reduce quality and diversity respectively, while not sensitive to others.
This indicates that they are valuable for diagnosing these individual failure modes but not for a rigorous evaluation or comparison. 
Diversity and coverage are also sensitive to these distributions, but their relationship to quality and diversity is less clear.
For example, the coverage is lower with the covariances multiplied by 10, when, in fact, the diversity should remain unchanged.
We, therefore, conclude that precision and recall are the more meaningful metrics to disentangle quality and diversity, and use those going forward.

\section{\label{sec:jetdata} Experiments on Jet Data}

We next test the performance of the Wasserstein distance, \fgdinf, MMD, precision, and recall on a realistic high energy physics dataset comprised of high momentum gluon jets.
As discussed in Sec.~\ref{sec:feature_selection}, we test all metrics on two sets of features per jet: (i) physically meaningful high-level features and (ii) features derived from a pretrained classifier.
We choose a set of 36 energy flow polynomials (EFPs)~\cite{komiske_efps} (all EFPs of degree less than five) for the former, as they form a complete basis for all infrared- and collinear-safe observables.
The classifier features are derived from the activations of the penultimate layer of the SOTA ParticleNet~\cite{qu_particlenet} classifier, as described in Ref.~\cite{kansal_mpgan}.
Finally, we test the binary classifier metric as in Refs.~\cite{krause_caloflow, calochallenge} using both ParticleNet directly on the low-level jet features and a two-layer fully connected network (FCN) on the high-level EFPs.
We note that Refs.~\cite{krause_caloflow, calochallenge} do not provide a recipe for measuring the null distribution, instead relying on direct comparisons between area under the curve (AUC) values, which is a limitation of this classifier-based metric.
We describe the dataset and tested distortions in Sec.~\ref{sec:jetdata_dataset}, and experimental results in Sec.~\ref{sec:jetdata_results}.

\subsection{\label{sec:jetdata_dataset} Dataset}

\begin{figure*}[ht]
    \includegraphics[width=\textwidth]{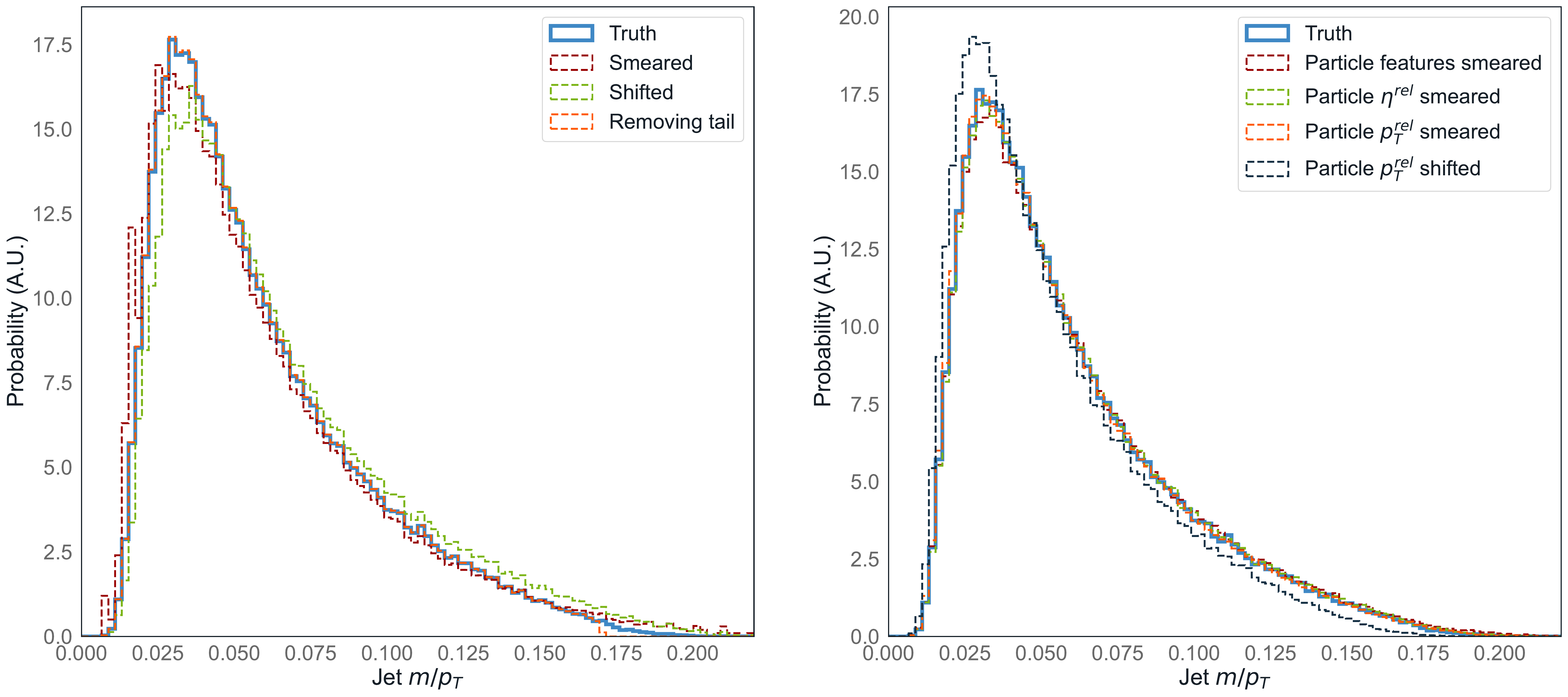}
    \caption{The probability, in arbitrary units (A.U.), of the relative jet mass for truth and distorted gluon jet distributions. 
    On the left are distribution-level distortions, and on the right particle-level.}
    \label{fig:jetdists} 
\end{figure*}

As our true distribution we use simulated gluon jets of ${\approx}1\TeV$ transverse momentum (\pt) from the \jetnet dataset~\cite{kansal_jetnet_dataset}, using the associated \jetnet library~\cite{kansal_jetnet_library}.
Details of the simulation and representation of the dataset can be found in Ref.~\cite{kansal_mpgan}.
We consider the three particle features: relative angular coordinates 
$\etarel =\eta^\mathrm{particle} - \eta^\mathrm{jet}$ and 
$\phirel =\phi^\mathrm{particle} - \phi^\mathrm{jet} \pmod{2\pi}$, and the relative transverse momentum 
$\ptrel = \pt^\mathrm{particle}/\pt^\mathrm{jet}$.
To obtain alternative distributions we distort the dataset in several ways typical of the mismodeling we observe in ML generative models: lower feature resolution, systematic shifts in the features, and inability to capture the full distribution.

We perform both distribution-level distortions, by reweighting the samples in jet mass to produce a mass distribution that is (i) smeared, (ii) smeared and shifted higher, and (iii) missing the tail of the distribution, as well as direct particle-level distortions, by (iv) smearing all three \phirel, \etarel, and \ptrel features, smearing the (v) \ptrel and (vi) \etarel individually, and (vii) shifting the \ptrel higher. 
The effects of the distortions on the relative jet mass are shown in Fig.~\ref{fig:jetdists}, with further plots of different variables available in Appendix~\ref{app:jet_plots}.

\subsection{\label{sec:jetdata_results} Results}

\begin{table*}[t]
\topcaption{Values, significances, and errors of metrics, as defined in Secs.~\ref{sec:toydata_metrics} and~\ref{sec:jetdata_results}, for each jet distribution, for the largest sample size tested. EFP and PN refer to metrics using EFPs and ParticleNet activations as their input features, respectively. 
The most significant scores per distribution are in bold.\label{tab:jet_results}}
\begin{adjustwidth}{-1in}{-1in}% adjust the L and R margins by 1 inch
\centering\resizebox{\textwidth}{!}{
    %\begin{tabular}{l|C{2cm}C{2cm}C{2cm}C{2cm}C{2cm}C{2cm}C{2cm}C{2cm}}
    \begin{tabular}{l|cccccccc}
    \toprule
    \input{tables/2_jet_measurements}\\
    \bottomrule
    \end{tabular}
}
\end{adjustwidth}
\end{table*}

Table~\ref{tab:jet_results} shows the central values, significances, and errors for each metric, as defined in Sec.~\ref{sec:toydata_metrics}, with the most significant scores per alternative distribution highlighted in bold.
The first row shows the Wasserstein distance between only the 1D jet mass distributions (\wassm) as introduced in Ref.~\cite{kansal_mpgan}, as a test of the power and limitations of considering only 1D marginal distributions.
% Using the same metric of significance as in Sec.~\ref{sec:toydata_results},
We see that, in fact, \wassm identifies most distortions as significantly discrepant but is not as sensitive to subtle changes such as particle \ptrel smearing.
Additionally, even with up to 50,000 samples, it is unable to converge to the true value.
Nevertheless, it proves to be a valuable metric that can be used for focused evaluation of specific physical features, complementing aggregate metrics.

The next five rows show values for metrics which use EFPs as their features.
We find that, perhaps surprisingly, \fgdinf is the most sensitive to all distortions, with significances orders of magnitude higher than the rest.
The Wasserstein distance is not sensitive to many distortions for the sample sizes tested, while the MMD is successful, but not as sensitive as \fgdinf.
It is also clear that precision and recall have difficulty discerning the quality and diversity of distributions in high-dimensional feature spaces, which is perhaps expected considering the difficulty of manifold estimation in such a space.

An extremely similar conclusion is reached when considering the metrics using ParticleNet activations, with \fgdinf again the highest performing.
Broadly, ParticleNet activations allow the metric to distinguish particle-level distortions slightly better, and vice versa for distribution-level distortions, although overall the sensitivities are quite similar.
% Considering the significant practical benefits of using hand-engineered features, in terms of ease of standardisation and interpretability, we conclude that they are the more appropriate set of features to use going forward.
We posit that including a subset of lower-level particle features in addition to EFPs could improve sensitivity to particle-level distortions, a study of which we leave to future work.

Finally, the last two rows provide the AUC values for a ParticleNet classifier trained on the particle low-level features (LLF), and an FCN trained on high-level features (HLF).
% We follow the prescription in Ref.~\cite{krause_caloflow}, and note 
We find that while both appear to be able to distinguish well the samples with particle-level distortions, they have no sensitivity to the distribution level distortions.

\begin{figure*}
    \includegraphics[width=\textwidth]{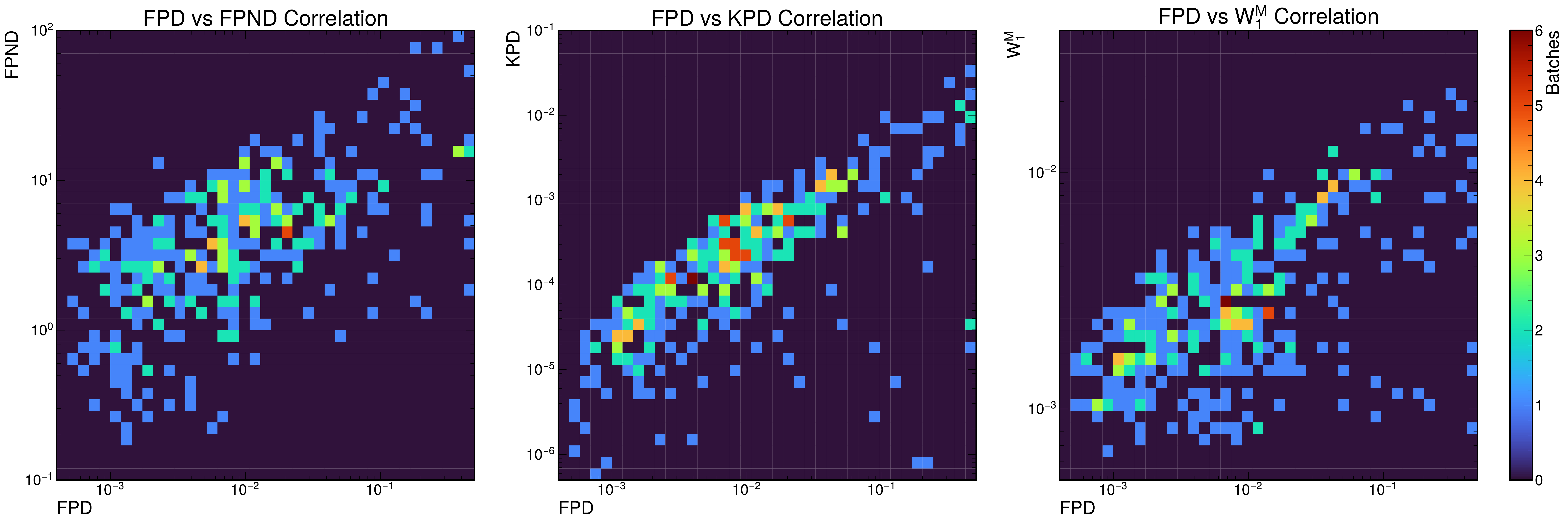}
    \caption{Correlations between FPD and FPND, KPD, and \wassm on 400 separate batches of 50,000 GAPT-generated jets.}
    \label{fig:correlations} 
\end{figure*}

In conclusion, we find from these experiments that \fgdinf is in fact the most sensitive metric to all distortions tested.
Despite the Gaussian assumption, it is clear that access to the first-order moments of the distribution is sufficient for it to have high discriminating power against the relevant alternative distributions we expect from generative models.

Applying \fgdinf to hand-engineered physical features or ParticleNet activations leads to similar performance, with the former having a slight edge.
In addition, \fgdinf using physical features---\textit{Fr\'echet physics distance (FPD)} for short---has a number of practical benefits.
For instance, it can be consistently applied to any data structure (e.g. point clouds or images) and easily adapted to different datasets as long as the same physical features can be derived from the data samples (Refs.~\cite{deOliveira:2017pjk, kansal_mpgan} derive similar jet observables from images and point clouds, respectively).
These are both difficult to do with features derived from a pretrained classifier, where different classifier architectures may need to be considered for different data structures and potentially even different datasets.
FPD is also more easily interpreted, as evaluators have more control and understanding of the set of features they provide as input.

Hence, we propose FPD as a novel efficient, interpretable, and highly sensitive metric for evaluating generative models in HEP.
However, MMD on hand-engineered features---kernel physics distance (KPD) for short ---and $W_1$ scores between individual feature distributions also provide valuable information and, as demonstrated in Sec.~\ref{sec:toydata}, can cover alternative distributions for which FPD lacks discriminating power.

\section{\label{sec:gapt} Demonstration on MPGAN versus GAPT}

\begin{figure*}[htpb]
    \includegraphics[width=\textwidth]{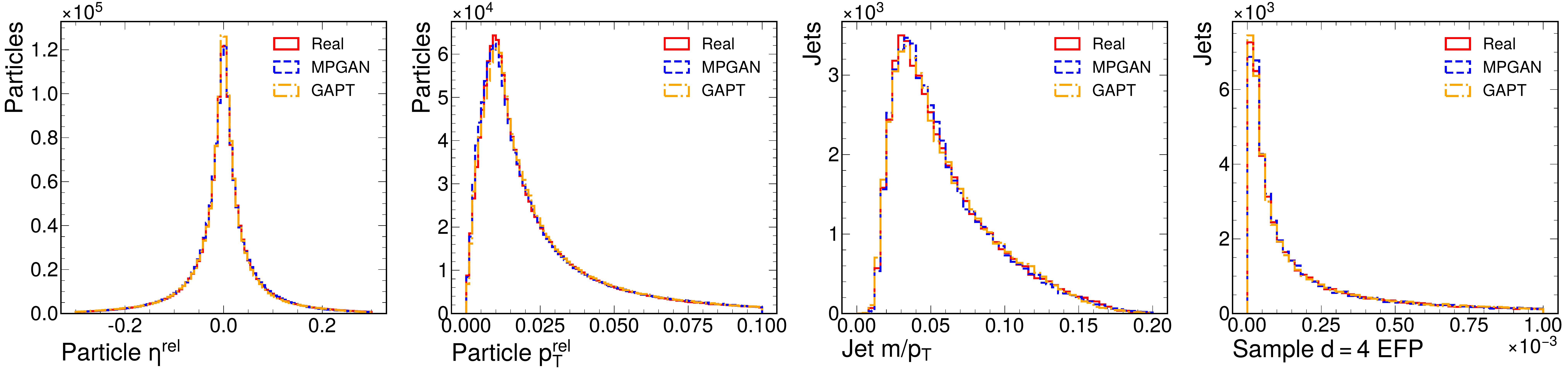}
    \caption{Low-level particle feature distributions (far left and center left) and high-level jet feature distributions (center right and far right) for the real data (red), MPGAN-generated data (blue), and GAPT-generated data (yellow).}
    \label{fig:featdists} 
\end{figure*}

\begin{table*}[htpb]
\topcaption{Values and errors of the proposed metrics on MPGAN and GAPT generated samples, as well as on separate samples from the true distribution.
The best scores per metric out of MPGAN and GAPT are highlighted in bold.
The inference time per jet is shown as well for both models.
\label{tab:modelscores}}
\centering{
    \begin{tabular}{l|C{2.5cm}C{2.5cm}C{2.5cm}C{2.5cm}C{2.5cm}}
    \toprule
    \input{tables/3_model_measurements}\\
    \bottomrule
    \end{tabular}
}
\end{table*}

We now provide a practical demonstration of the efficacy of our proposed metrics in evaluating two high-performing generative models: the message-passing GAN (MPGAN) and the new generative adversarial particle transformer (GAPT), on the same gluon jet dataset described in Sec.~\ref{sec:jetdata}.
We describe the models in detail in Sec.~\ref{sec:gapt_models}, and discuss experimental results in~\ref{sec:gapt_results}.

\subsection{\label{sec:gapt_models} MPGAN and GAPT}

MPGAN is a graph-based GAN using a message-passing framework for the generator and discriminator networks, which is the current SOTA in simulating particle clouds---jets represented as point clouds in momentum space.
We use the same architecture and pretrained models provided by Ref.~\cite{kansal_mpgan} to generate gluon jets from the \jetnet dataset.

We also introduce GAPT, a new GAN for particle cloud data which employs self-attention instead of message passing in the two networks.
It is based on the generative adversarial set transformer (GAST) architecture~\cite{stelzner_gast}, which makes use of set transformer~\cite{lee_settransformer} blocks to aggregate information across all points and update their features. 
It maintains the key inductive biases which makes MPGAN successful---permutation symmetry-respecting operations, and fully connected interaction between nodes during generation to learn high-level global features, but with a significant improvement in speed (Table~\ref{tab:modelscores}), and promising avenues to scale the architecture linearly in the number of nodes using induced self-attention blocks~\cite{lee_settransformer} and conditional generation~\cite{stelzner_gast}.
The code for both models is available in Ref.~\cite{kansal_mpgan_code}.
Further training and implementation details can be found in Appendix~\ref{app:training}.

Because of the similarity in architectures of MPGAN and GAPT, both visually and using the 1D $W_1$ physics-based metrics discussed in Ref.~\cite{kansal_mpgan}, it is difficult to discern which is more performant. 
Hence, this makes an effective test bench for our proposed metrics. 

\subsection{\label{sec:gapt_results} Results}

Figure~\ref{fig:correlations} shows correlation plots between FPD and FPND, KPD, and \wassm on 400 separate batches of 50,000 GAPT-generated jets.
We observe an overall positive relationship between the metrics, as one might expect. 
FPD and KPD have the strongest correlation, likely because they are accessing similar information about the same set of input features.
However, for low values, the correlation is weak between all metrics, indicating that these metrics are complementary in understanding different aspects of the model's performance.
As noted in Sec.~\ref{sec:jetdata_results}, the correlation between FPD and FPND may improve if the former were to use a subset of lower-level particle features as well.

Histograms of sample feature distributions and FPD, KPD, \wassm scores as well as the $W_1$ distance between the particle \ptrel distributions from the best-performing MPGAN, as provided by Ref.~\cite{kansal_mpgan}, and GAPT, based on FPD, models are shown in Fig.~\ref{fig:featdists} and Table~\ref{tab:modelscores}, respectively.
For completeness, Table~\ref{tab:modelscores} also shows measurements of the inference time per jet for each model, measured on an NVIDIA RTX A6000 GPU.
It is extremely difficult to either distinguish between the performance of the two models or draw a conclusion for their viability as alternative simulators based only on visual inspection of the histograms or even the 1D $W_1$ scores.
However, FPD and KPD provide crucial information in this regard, with FPD more sensitive as expected from Sec.~\ref{sec:jetdata}, clearly indicating that MPGAN significantly outperforms GAPT, but its samples remain discrepant from the true distribution.
We note, however, that despite the suboptimal performance, GAPT provides the benefits of speed and scalability.

Overall, we see from this experiment the value in employing a broad set of sensitive, interpretable metrics.
Firstly, evaluators can identify specific points of failures in their models.
In the case of GAPT, we note that while its \wassm and FPD scores are significantly discrepant from the truth, the \wassppt score is consistent with the truth within the statistical uncertainty, implying that GAPT is modeling the particle features accurately but work is needed on its modeling of interparticle correlations.
Secondly, evaluators are also able to define clear, quantitative criteria for model selection for their downstream tasks: for example, if comparing different simulator options, they can simply choose the model with the lowest FPD score, or if validating a faster alternative to traditional, accurate simulations, they may wish to require all scores to be compatible (e.g., significances of $<2$) with the latter, or even with LHC data itself, before adopting the model.

\section{\label{sec:conclusion} Conclusion}

We have discussed several potential evaluation metrics for generative models in HEP, using the framework of two-sample GOF testing between real and simulated data.
Inspired by validation of simulations in both physics and machine learning, we introduce two new metrics, the Fr\'echet and kernel physics distances, which employ hand-engineered physical features, to compare and evaluate alternative simulators.
Practically, these metrics are efficient, reproducible, and easily standardized, and, being multivariate, can be naturally extended to conditional generation.

We performed a variety of experiments using the proposed metrics on toy Gaussian-distributed and high energy jet data.
We illustrated as well the power of these metrics to discern between two state-of-the-art ML models for simulating jets: the MPGAN and our newly developed GAPT, which is significantly faster.
We find that FPD is extremely sensitive to expected distortions from ML generative models, and collectively, FPD, KPD and the Wasserstein 1-distance ($W_1$) between individual feature distributions, should successfully cover all relevant alternative generated distributions.
Hence, we recommend the adoption of these metrics in HEP for evaluating generative models. 
Future work may explore the specific set of physical features for jets, calorimeter showers, and beyond, to use for FPD and KPD.

\begin{acknowledgments}
%%IRIS-HEP fellowship acknowledgment
R.~K. and A.~L were partially supported by an Institute for Research and Innovation in Software for High Energy Physics (IRIS-HEP) fellowship through the U.S. National Science Foundation (NSF) under Cooperative Agreement No. OAC-1836650.
R.~K. was additionally supported by the LHC Physics Center at Fermi National Accelerator Laboratory, managed and operated by Fermi Research Alliance, LLC under Contract No. DE-AC02-07CH11359 with the U.S. Department of Energy (DOE).
J.~D. is supported by the DOE, Office of Science, Office of High Energy Physics Early Career Research program under Grant No. DE-SC0021187, the DOE, Office of Advanced Scientific Computing Research under Grant No. DE-SC0021396, Findable, Accessible, Interoperable, and Reusable Frameworks for Physics-Inspired Artificial Intelligence in High Energy Physics (FAIR4HEP), and the NSF Harnessing the Data Revolution (HDR) Institute for Accelerating AI Algorithms for Data Driven Discovery (A3D3) under Cooperative Agreement No. OAC-2117997.
M.~P. and N.~C. were supported by the European Research Council (ERC) under the European Union's Horizon 2020 research and innovation program (Grant Agreement No. 772369).
B.~O and T.~T are supported by Grant No. 2018/25225-9, S\~{a}o Paulo Research Foundation (FAPESP).
B.~O was also partially supported by Grants No. 2019/16401-0 and No. 2020/06600-3, S\~{a}o Paulo Research Foundation (FAPESP).
This work was performed using the Pacific Research Platform Nautilus HyperCluster supported by NSF Grants No. CNS-1730158, No. ACI-1540112, No. ACI-1541349, No. OAC-1826967, the University of California Office of the President, and the University of California San Diego's California Institute for Telecommunications and Information Technology/Qualcomm Institute. 
Thanks to CENIC for the 100\,Gpbs networks.
Funding for cloud credits was supported by NSF Grant No. 1904444  Internet2 supported E-CAS Exploring Clouds to Accelerate Science.
\end{acknowledgments}

%\clearpage
% \bibliographystyle{cms_unsrt}
% \bibliographystyle{apsrev4-2}
\bibliography{bibliography} % Produces the bibliography via BibTeX.

\clearpage
\appendix

\section{Further Discussion on IPMs vs. \texorpdfstring{$f$}{f}-Divergences}
\label{app:metricspace} 

\begin{figure*}[htpb]
    \includegraphics[width=\textwidth]{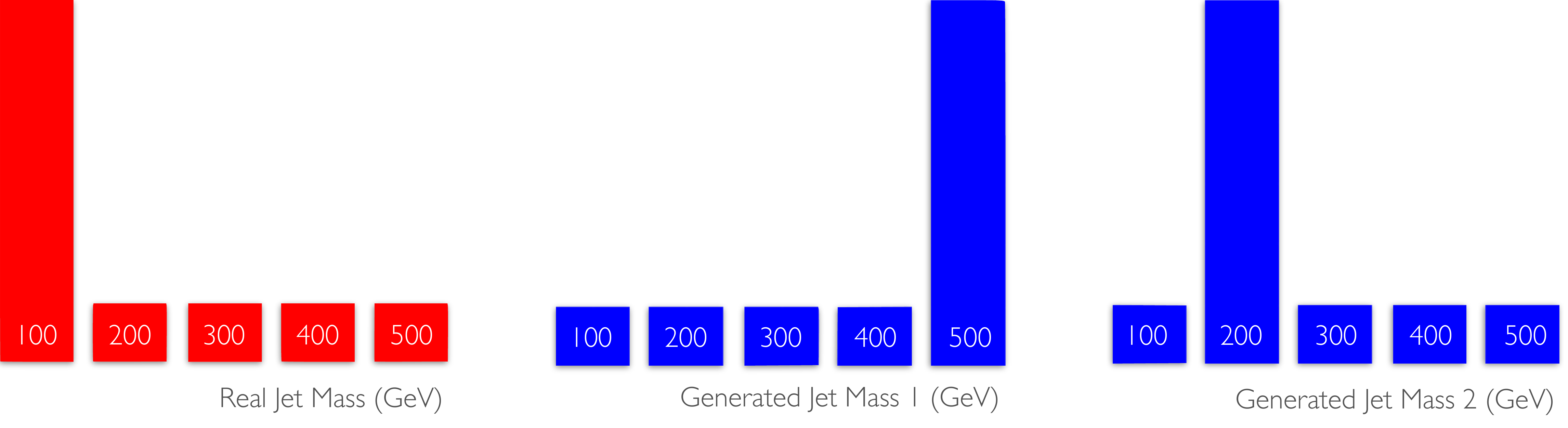}
    \caption{Example real and generated jet mass distributions used to illustrate the benefit of IPMs in Appendix~\ref{app:metricspace}, based on Refs.~\cite{gretton_talk, w1_stackoverflow}.}
    \label{fig:metricspace}
\end{figure*}

A crucial advantage of IPMs in evaluating generative models is that they consider the metric space of the distributions.
We illustrate this with the help of Fig.~\ref{fig:metricspace}, inspired heavily by Refs.~\cite{gretton_talk, w1_stackoverflow}, which shows an example real (in red) and two generated (in blue) jet mass distributions.
Clearly, in the context of simulation, the second generated distribution contains a peak closer to the real peak and, hence, is a better model.
However, because $f$-divergences such as the KL or $\chi^2$ look only at the pointwise difference between distributions, they find both generated distributions to be as discrepant with the real.
IPMs like the Wasserstein metric or MMD, on the other hand, generally identify the second distribution as being closer to the real.

\section{\label{app:details} Further Discussion on Gaussian Dataset Experiments}

\begin{figure*}[htpb]
    \includegraphics[width=\textwidth]{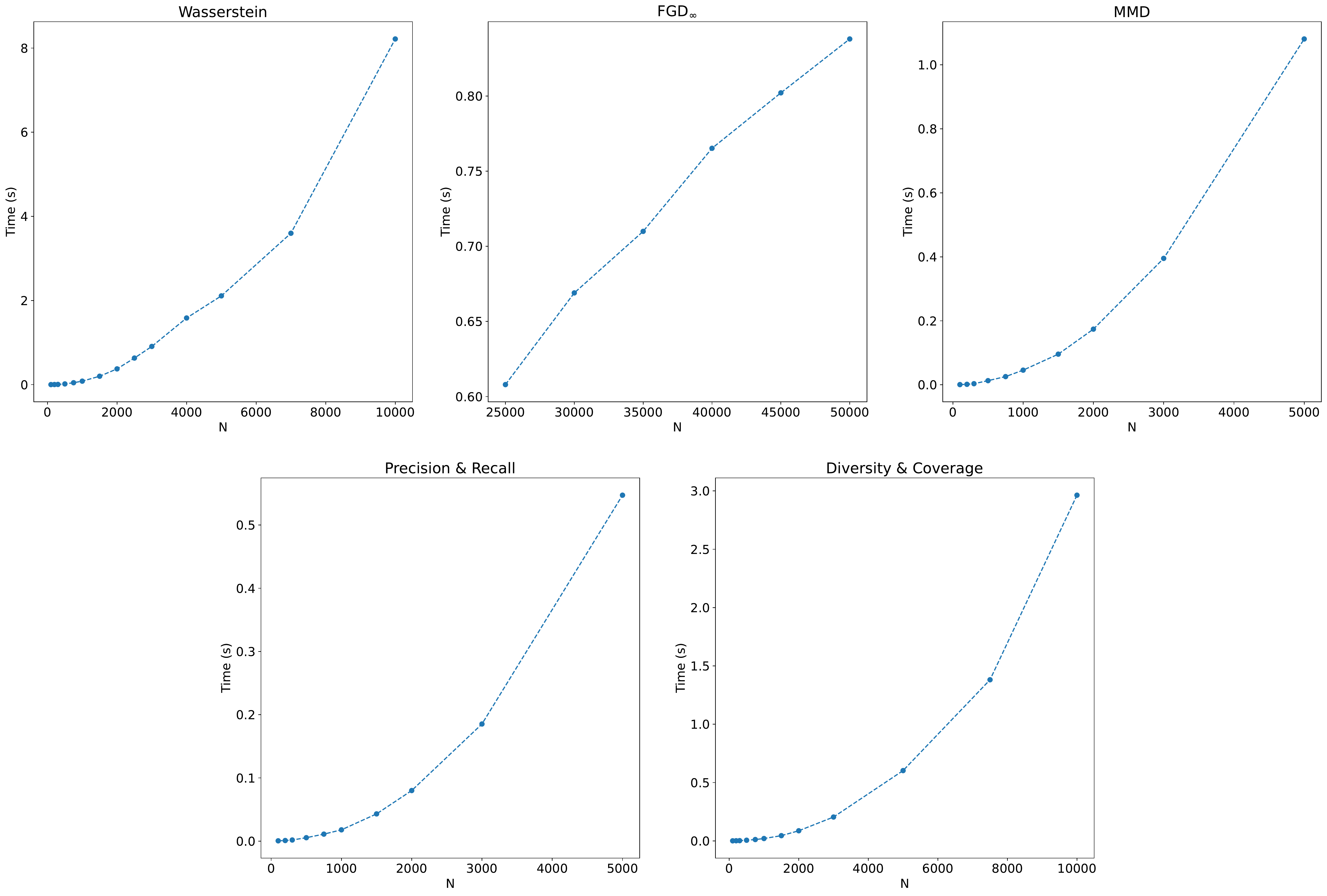}
    \caption{Time taken per each metric on Gaussian-distributed datasets as described in Sec.~\ref{sec:toydata}.}
    \label{fig:timings}
\end{figure*}

\begin{figure*}[htpb]
    \includegraphics[width=0.75\textwidth]{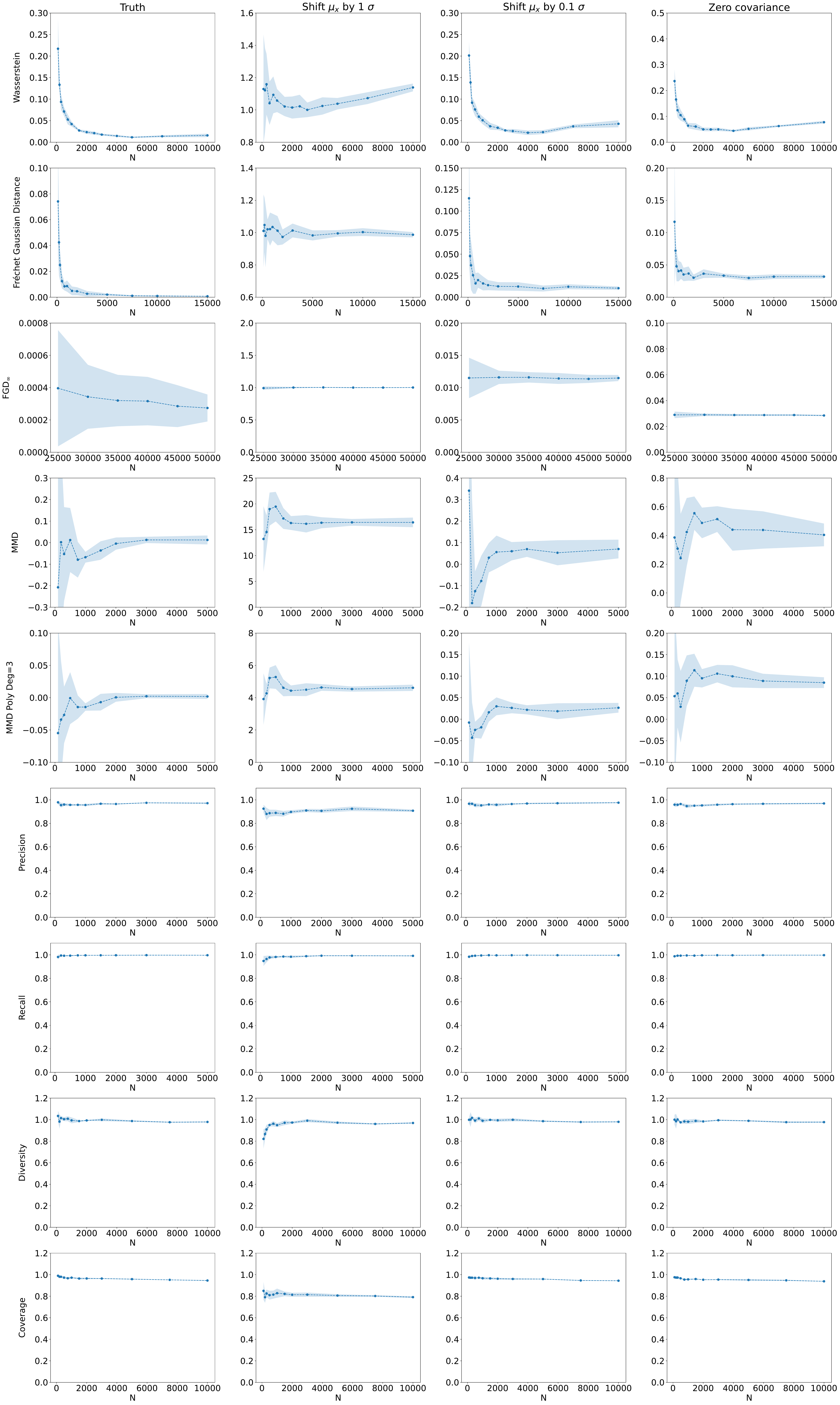}
    \caption{Scores of each metric on Gaussian-distributed datasets as described in Sec.~\ref{sec:toydata}.}
    \label{fig:toyscores1}
\end{figure*}

\begin{figure*}[htpb]
    \includegraphics[width=0.75\textwidth]{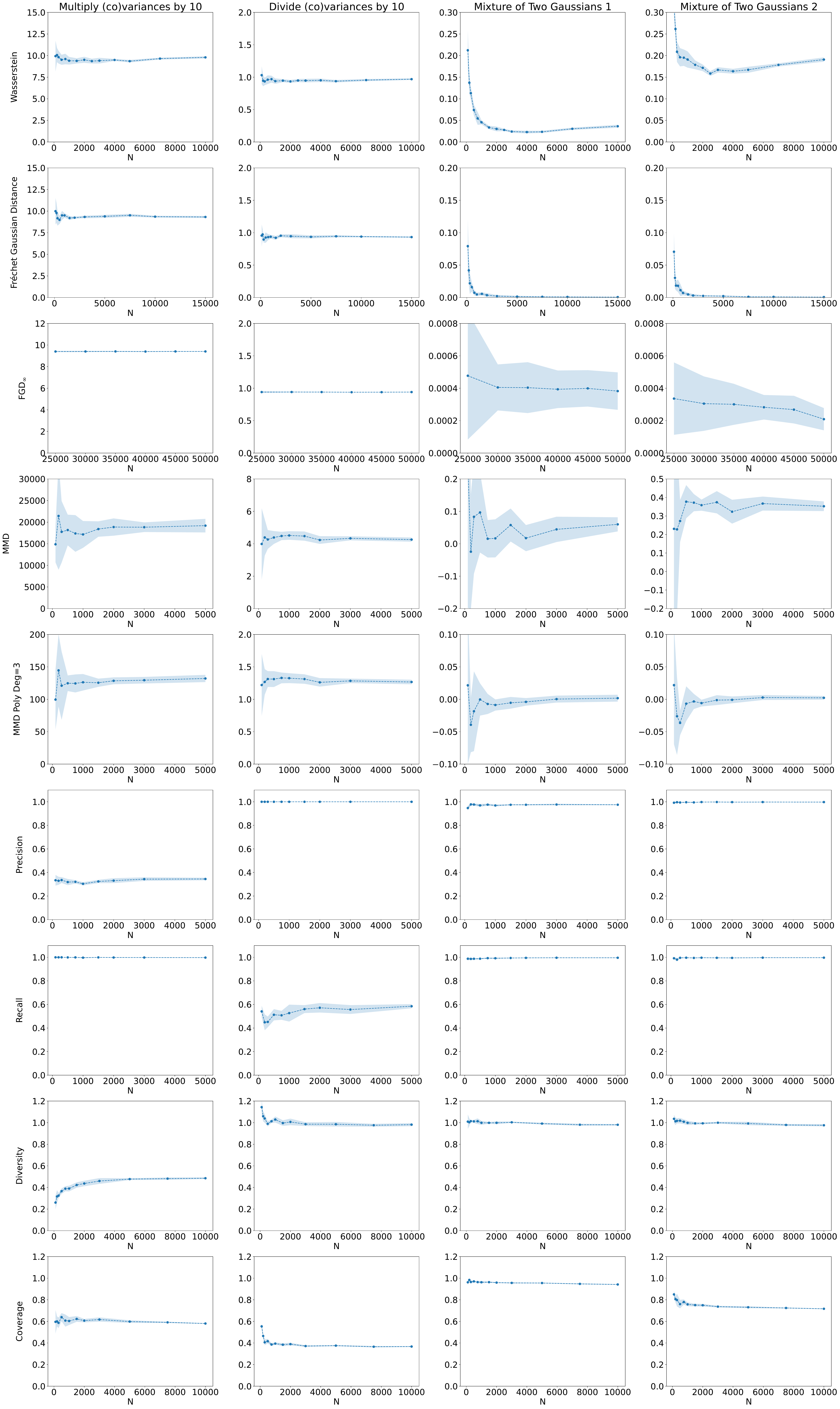}
    \caption{Scores of each metric on Gaussian-distributed datasets as described in Sec.~\ref{sec:toydata}.}
    \label{fig:toyscores2}
\end{figure*}

Figure~\ref{fig:timings} plots the time taken per measurement of each metric used in Sec.~\ref{sec:toydata} for different sample sizes, measured on an 8-core Intel Core i9 processor.
The quadratic scaling of the Wasserstein and diversity and coverage metrics, in combination with their low rate of convergence, means their use for evaluation is practically difficult.
MMD and precision and recall exhibit the same scaling; however, are observed to converge within roughly 3000 samples.
\fgdinf scales linearly and remains fast to compute even at the highest batch size tested.

Figures~\ref{fig:toyscores1} and~\ref{fig:toyscores2} show measurements of each metric on each distribution discussed in Sec.~\ref{sec:toydata}, as well as FGD and MMD with a third-order polynomial kernel for varying samples sizes.
We can see from these plots that indeed, as discussed in Refs.~\cite{binkowski_demystifying, chong_unbiasedfid}, FGD is biased, but the solution from Ref.~\cite{chong_unbiasedfid} of extrapolating to infinite-sample size (\fgdinf) largely solves this issue.
We also note that, perhaps surprisingly, a third-order polynomial kernel, as used for the KID~\cite{binkowski_demystifying} in computer vision, is not sufficient to discern the mixtures of Gaussian distributions from the single Gaussian.
Hence, we recommend a fourth-order kernel for the kernel physics distance.

\section{\label{app:jet_plots} Alternative Jet Distributions}

\begin{figure*}[htpb]
    \includegraphics[width=0.94\textwidth]{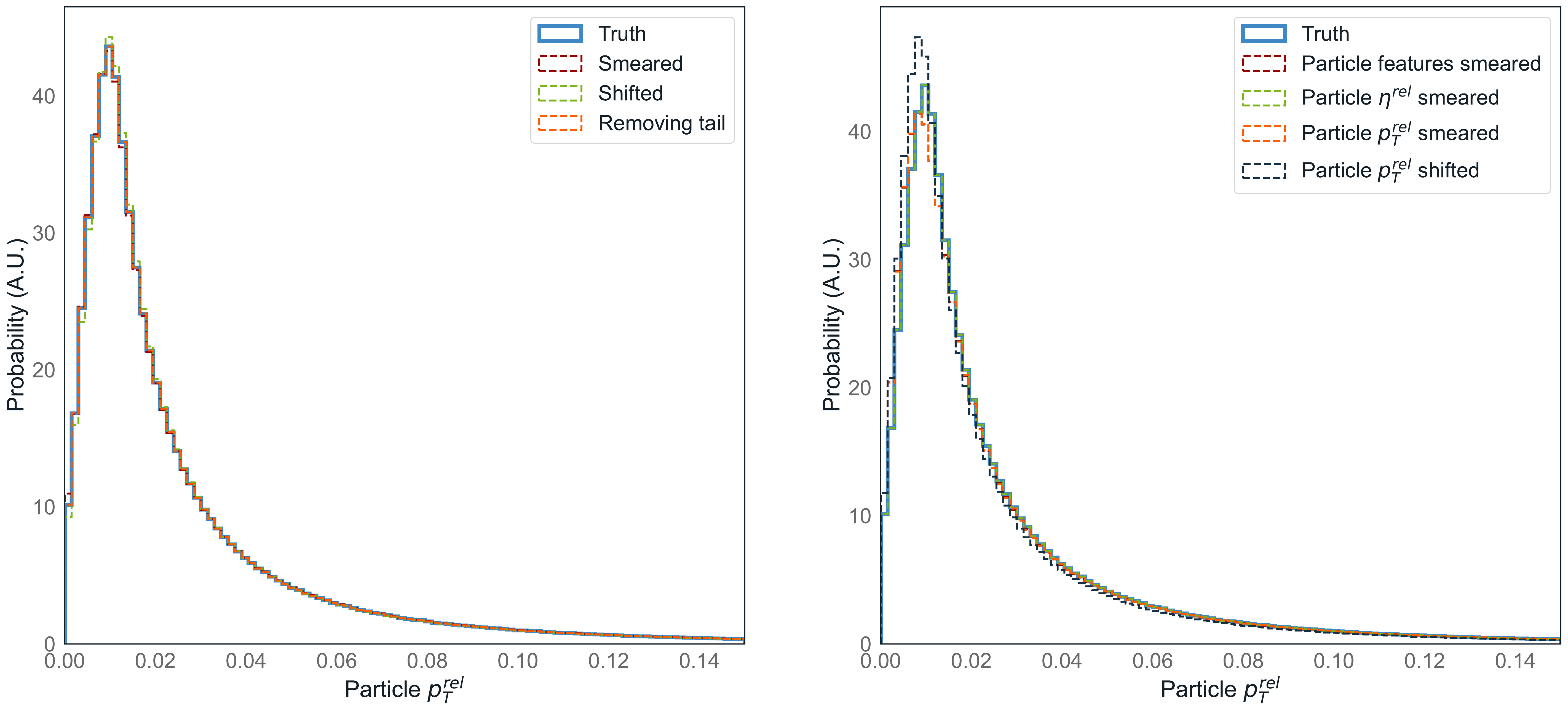}
    \includegraphics[width=0.94\textwidth]{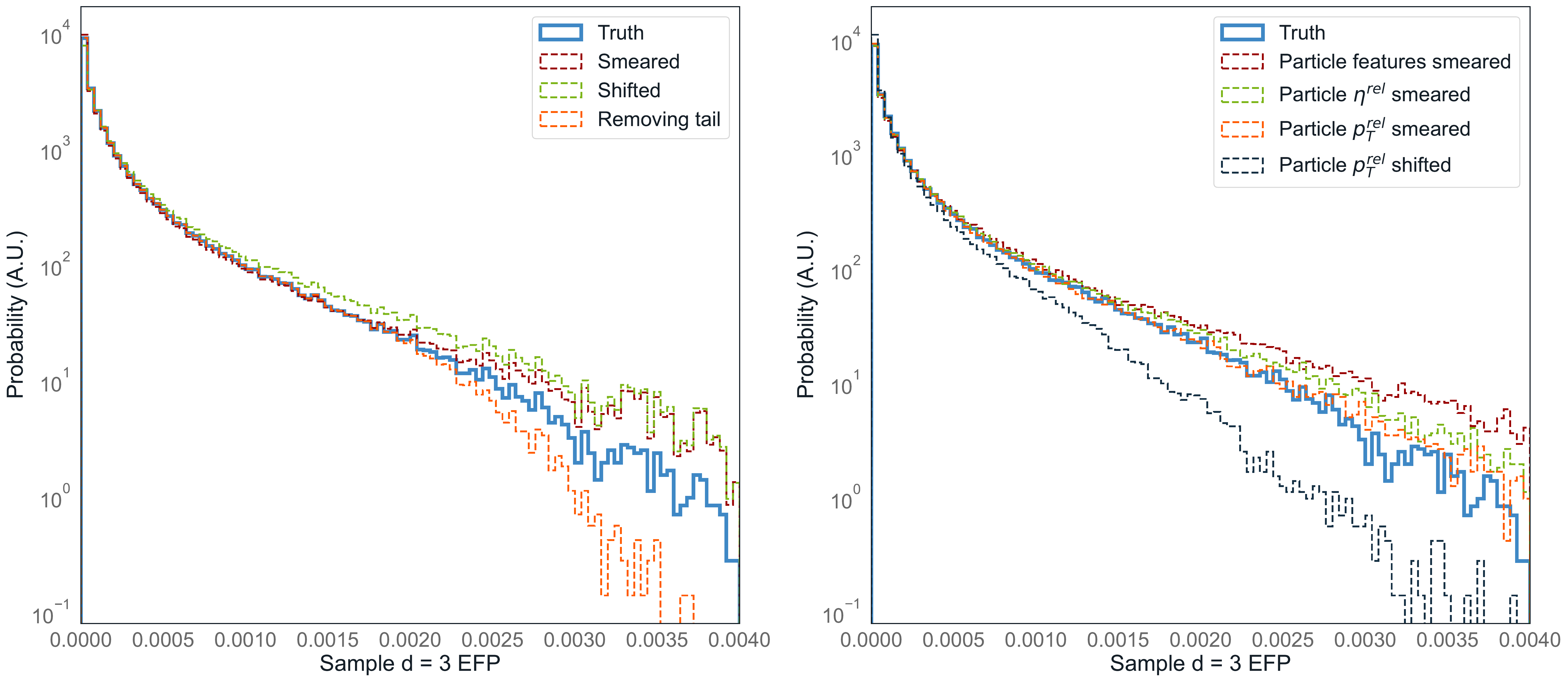}
    \includegraphics[width=0.94\textwidth]{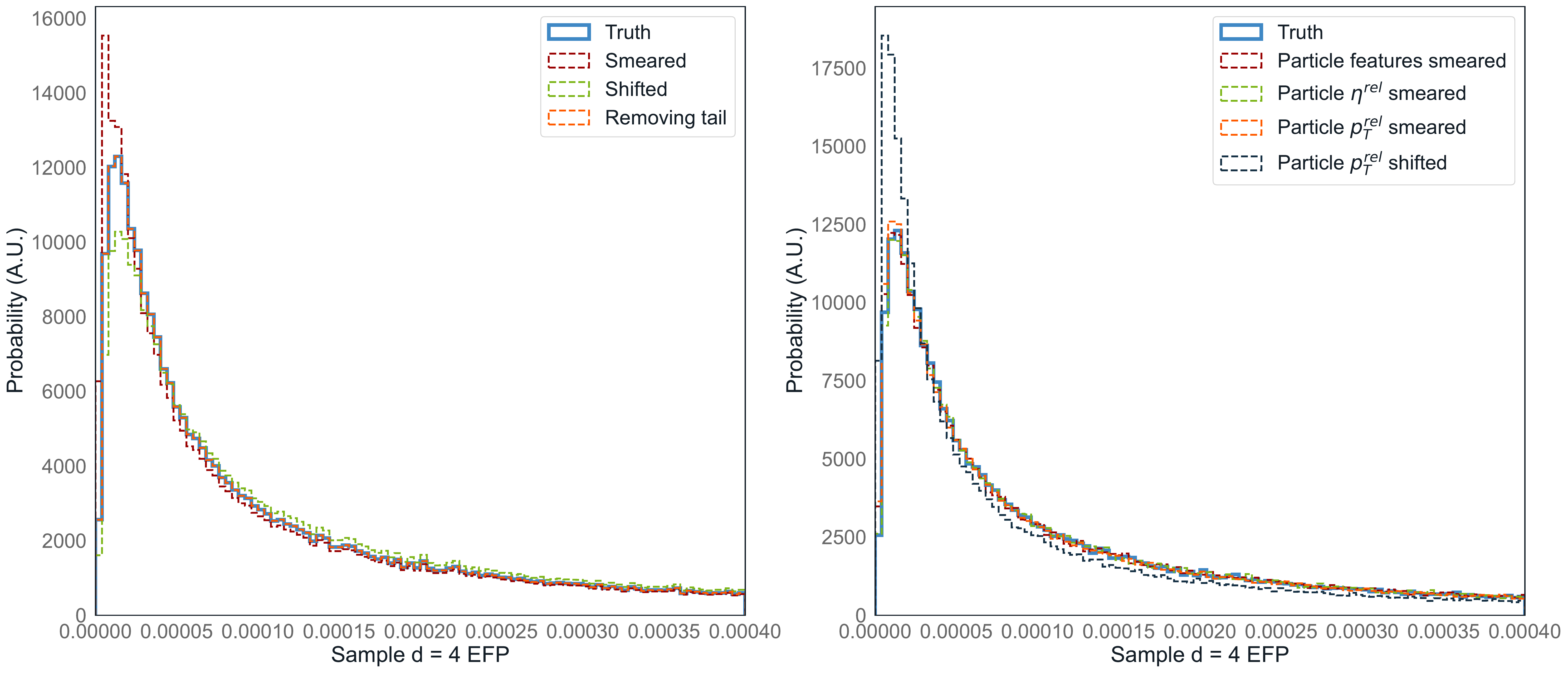}
    \caption{The probability, in arbitrary units (A.U.), of the particle \ptrel, a sample $d=3$ EFP, and a sample $d=4$ EFP for truth and distorted gluon jet distributions. 
    On the left are distribution-level distortions, and on the right particle-level.}
    \label{fig:jet_dists_app}
\end{figure*}

Distributions of particle- and jet-level features from the true and distorted jets as described in Sec. \ref{sec:jetdata} are shown in Fig.~\ref{fig:jet_dists_app}.\\

\section{\label{app:training} Neural Network Implementation and Training Details}

\subsection{Binary classifiers}

In Sec.~\ref{sec:jetdata}, we use the ParticleNet~\cite{qu_particlenet} graph neural network model and a simple dense fully-connected neural network as binary classifiers on low-level particle and high-level features, respectively, for evaluating true and distorted jet distributions.
We use the same architecture and hyperparameters for ParticleNet as described in Ref.~\cite{qu_particlenet}.
We use two hidden layers of 128 neurons each with leaky rectified linear unit (ReLU) activations, with negative slope coefficient 0.2, after all but the final layer, which uses a sigmoid activation function.
We follow the same training scheme as in Ref.~\cite{qu_particlenet}, with a 70\%/30\% training/validation split on the dataset, and report the highest AUC achieved by each model on the validation dataset in Table~\ref{tab:jet_results}.

\subsection{GAPT}

GAPT is based on the GAST model~\cite{stelzner_gast}.
The generator and discriminator networks are composed of permutation-invariant multihead self-attention blocks (SABs) as defined in Ref.~\cite{lee_settransformer}.
We use four and two SAB blocks in the generator and discriminator respectively.
Each SAB block uses 8 attention heads, and a 128-dimensional embedding space for each of the query, key, and value vectors.
It also contains a one-layer feed-forward neural network (FFN) after each attention step, which maintains the 128-dimensional embedding for node features, and applies a leaky ReLU activation, with negative slope coefficient 0.2.
Residual connections to the pre-SAB node features are used after both the attention step and FFN.
After the final SAB block, a tanh activation is applied to the generator, whereas in the discriminator, the results are first pooled using a pooling by multihead attention (PMA) block~\cite{lee_settransformer}, followed by a final fully connected layer and sigmoid activation.

For training, we use the mean squared error loss function, as in the LSGAN~\cite{mao_lsgan}, and the RMSProp optimizer with a two timescale update rule~\cite{TTUR}, using a learning rate of $3\cdot10^{-4}$ and $10^{-4}$ for the discriminator and generator respectively.
Dropout, with probability 0.5, is used to regularize the discriminator.
We train for 2000 epochs and select the model with the lowest Fr\'echet physics distance.

\end{document}

%% file: tables/1_toy_measurements.tex
\multirow{3}{*}{Metric} & \multirow{3}{*}{Truth} & \multirow{3}{*}{\vspace{2mm}\parbox{2cm}{Shift $\mu_x$ by 1$\sigma$}} & \multirow{3}{*}{\vspace{2mm}\parbox{2cm}{Shift $\mu_x$ by 0.1$\sigma$}} & \multirow{3}{*}{\vspace{2mm}\parbox{2cm}{Zero covariance}} & Multiply (co)variances by 10 & Divide (co)variances by 10 & Mixture of Two Gaussians 1 & Mixture of Two Gaussians 2\\ \midrule
Wasserstein & $0.016 \pm 0.004$ & $1.14 \pm 0.02$ & $0.043 \pm 0.008$ & $0.077 \pm 0.006$ & $9.8 \pm 0.1$ & $0.97 \pm 0.01$ & $\mathbf{0.036 \pm 0.003}$ & $0.191 \pm 0.005$\\
Significance & $ $ & $284 \pm 6$ & $7 \pm 1$ & $16 \pm 1$ & $2460 \pm 30$ & $241 \pm 3$ & $\mathbf{5.2 \pm 0.5}$ & $44 \pm 1$\\ \midrule
$\mathrm{FGD}_\infty$ $\times 10^3$ & $0.27 \pm 0.08$ & $\mathbf{1002 \pm 4}$ & $\mathbf{11.5 \pm 0.5}$ & $\mathbf{28.4 \pm 0.5}$ & $9400 \pm 20$ & $\mathbf{941 \pm 2}$ & $0.4 \pm 0.1$ & $0.21 \pm 0.07$\\
Significance & $ $ & $\mathbf{11960 \pm 40}$ & $\mathbf{134 \pm 6}$ & $\mathbf{336 \pm 6}$ & $112300 \pm 200$ & $\mathbf{11230 \pm 20}$ & $1.3 \pm 0.4$ & 0\\ \midrule 
MMD & $0.01 \pm 0.02$ & $16.4 \pm 0.9$ & $0.07 \pm 0.04$ & $0.40 \pm 0.08$ & $\mathbf{19\mathrm{k} \pm 1\mathrm{k}}$ & $4.3 \pm 0.1$ & $0.06 \pm 0.02$ & $0.35 \pm 0.03$\\
Significance & \ & $790 \pm 40$ & $3 \pm 2$ & $19 \pm 4$ & $\mathbf{920\mathrm{k} \pm 70\mathrm{k}}$ & $204 \pm 6$ & $2.3 \pm 0.8$ & $16 \pm 1$\\ \midrule 
Precision & $0.972 \pm 0.005$ & $0.91 \pm 0.01$ & $0.976 \pm 0.004$ & $0.969 \pm 0.006$ & $0.34 \pm 0.01$ & $1.0 \pm 0.0$ & $0.975 \pm 0.003$ & $0.998 \pm 0.001$\\
Significance & \ & $12.2 \pm 0.1$ & 0 & $0.440 \pm 0.003$ & $119 \pm 4$ & 0 & 0 & 0\\ \midrule 
Recall & $0.997 \pm 0.001$ & $0.992 \pm 0.003$ & $0.997 \pm 0.001$ & $0.998 \pm 0.001$ & $0.998 \pm 0.001$ & $0.58 \pm 0.02$ & $0.996 \pm 0.001$ & $0.997 \pm 0.001$\\
Significance & \ & $5.38 \pm 0.02$ & $0.227 \pm 0.000$ & 0 & 0 & $420 \pm 10$ & $0.762 \pm 0.001$ & 0\\ \midrule 
Diversity & $0.979 \pm 0.005$ & $0.969 \pm 0.007$ & $0.980 \pm 0.005$ & $0.977 \pm 0.006$ & $0.486 \pm 0.007$ & $0.98 \pm 0.01$ & $0.981 \pm 0.007$ & $0.98 \pm 0.01$\\
Significance & \ & $2.11 \pm 0.02$ & 0 & $0.335 \pm 0.002$ & $109 \pm 2$ & 0 & 0 & $0.654 \pm 0.007$\\ \midrule 
Coverage & $0.946 \pm 0.004$ & $0.791 \pm 0.008$ & $0.944 \pm 0.002$ & $0.939 \pm 0.002$ & $0.580 \pm 0.003$ & $0.367 \pm 0.004$ & $0.942 \pm 0.003$ & $\mathbf{0.717 \pm 0.003}$\\
Significance & \ & $43.8 \pm 0.4$ & $0.493 \pm 0.001$ & $2.047 \pm 0.004$ & $103.6 \pm 0.6$ & $164 \pm 2$ & $1.094 \pm 0.004$ & $\mathbf{64.9 \pm 0.3}$

%% file: tables/2_jet_measurements.tex
\multirow{3}{*}{Metric} & \multirow{3}{*}{Truth} & \multirow{3}{*}{Smeared} & \multirow{3}{*}{Shifted} & \multirow{3}{*}{Removing tail} & Particle & Particle & Particle & Particle \\
 & & & & & features & $\eta^\mathrm{rel}$ & $\pt^\mathrm{rel}$ & $\pt^\mathrm{rel}$\\
 & & & & & smeared & smeared & smeared & shifted \\\midrule
$W_1^M \times 10^3$ & $0.28 \pm 0.05$ & $2.1 \pm 0.2$ & $6.0 \pm 0.3$ & $0.6 \pm 0.2$ & $1.7 \pm 0.2$ & $0.9 \pm 0.3$ & $0.5 \pm 0.2$ & $5.8 \pm 0.2$\\
Significance &  & $37 \pm 3$ & $114 \pm 6$ & $7 \pm 2$ & $28 \pm 3$ & $12 \pm 4$ & $4 \pm 1$ & $111 \pm 3$\\ \midrule  \midrule 
Wasserstein EFP & $0.02 \pm 0.01$ & $0.09 \pm 0.05$ & $0.10 \pm 0.02$ & $0.016 \pm 0.007$ & $0.19 \pm 0.08$ & $0.03 \pm 0.01$ & $0.03 \pm 0.02$ & $0.06 \pm 0.02$\\
Significance &  & $6 \pm 4$ & $7 \pm 1$ & $0.06 \pm 0.02$ & $14 \pm 6$ & $0.8 \pm 0.4$ & $0.9 \pm 0.6$ & $4 \pm 1$\\ \midrule 
$\mathrm{FGD}_{\infty}$ EFP $\times 10^3$ & $0.08 \pm 0.03$ & $\mathbf{20 \pm 1}$ & $\mathbf{26.6 \pm 0.9}$ & $\mathbf{2.4 \pm 0.1}$ & $21 \pm 2$ & $\mathbf{3.6 \pm 0.3}$ & $2.3 \pm 0.2$ & $29.1 \pm 0.4$\\
Significance &  & $\mathbf{580 \pm 30}$ & $\mathbf{760 \pm 20}$ & $\mathbf{66 \pm 4}$ & $610 \pm 40$ & $\mathbf{103 \pm 8}$ & $64 \pm 4$ & $830 \pm 10$\\ \midrule 
MMD EFP $\times 10^3$ & $-0.006 \pm 0.005$ & $0.17 \pm 0.06$ & $0.9 \pm 0.1$ & $0.03 \pm 0.02$ & $0.35 \pm 0.09$ & $0.08 \pm 0.05$ & $0.01 \pm 0.02$ & $1.8 \pm 0.1$\\
Significance &  & $30 \pm 10$ & $170 \pm 20$ & $6 \pm 4$ & $70 \pm 10$ & $10 \pm 10$ & $3 \pm 5$ & $360 \pm 20$\\ \midrule 
Precision EFP & $0.9 \pm 0.1$ & $0.94 \pm 0.04$ & $0.978 \pm 0.005$ & $0.88 \pm 0.08$ & $0.7 \pm 0.1$ & $0.94 \pm 0.06$ & $0.7 \pm 0.1$ & $0.79 \pm 0.09$\\
Significance &  & 0 & 0 & $0.109 \pm 0.009$ & $1.9 \pm 0.3$ & 0 & $2.0 \pm 0.3$ & $0.9 \pm 0.1$\\ \midrule 
Recall EFP & $0.9 \pm 0.1$ & $0.88 \pm 0.07$ & $0.97 \pm 0.01$ & $0.92 \pm 0.06$ & $0.83 \pm 0.05$ & $0.92 \pm 0.07$ & $0.8 \pm 0.1$ & $0.8 \pm 0.1$\\
Significance &  & $0.16 \pm 0.01$ & 0 & 0 & $0.58 \pm 0.04$ & 0 & $0.8 \pm 0.1$ & $1.1 \pm 0.2$\\ \midrule \midrule 
Wasserstein PN & $1.65 \pm 0.06$ & $1.7 \pm 0.1$ & $2.4 \pm 0.4$ & $1.71 \pm 0.08$ & $4.5 \pm 0.1$ & $1.79 \pm 0.05$ & $4.0 \pm 0.4$ & $7.6 \pm 0.2$\\
Significance &  & $0.84 \pm 0.05$ & $12 \pm 2$ & $0.97 \pm 0.05$ & $45 \pm 1$ & $2.26 \pm 0.06$ & $37 \pm 3$ & $95 \pm 3$\\ \midrule 
$\mathrm{FGD}_{\infty}$ PN $\times 10^3$ & $0.6 \pm 0.4$ & $37 \pm 2$ & $202 \pm 4$ & $4.3 \pm 0.4$ & $\mathbf{1220 \pm 10}$ & $20 \pm 1$ & $\mathbf{1230 \pm 10}$ & $\mathbf{3630 \pm 10}$\\
Significance &  & $98 \pm 4$ & $540 \pm 0$ & $9.8 \pm 0.9$ & $\mathbf{3320 \pm 20}$ & $51 \pm 3$ & $\mathbf{3340 \pm 30}$ & $\mathbf{9870 \pm 30}$\\ \midrule 
MMD PN $\times 10^3$ & $-2 \pm 2$ & $4 \pm 8$ & $80 \pm 10$ & $-1 \pm 4$ & $500 \pm 100$ & $3 \pm 2$ & $560 \pm 60$ & $1100 \pm 40$\\
Significance &  & $3 \pm 6$ & $40 \pm 10$ & $0 \pm 3$ & $280 \pm 70$ & $3 \pm 2$ & $310 \pm 30$ & $610 \pm 20$\\ \midrule 
Precision PN & $0.68 \pm 0.07$ & $0.64 \pm 0.04$ & $0.71 \pm 0.06$ & $0.73 \pm 0.03$ & $0.09 \pm 0.04$ & $0.75 \pm 0.08$ & $0.08 \pm 0.04$ & $0.39 \pm 0.08$\\
Significance &  & $0.57 \pm 0.04$ & 0 & 0 & $8 \pm 4$ & 0 & $8 \pm 5$ & $4.0 \pm 0.8$\\ \midrule 
Recall PN & $0.70 \pm 0.05$ & $0.61 \pm 0.04$ & $0.61 \pm 0.08$ & $0.73 \pm 0.06$ & $0.014 \pm 0.009$ & $0.7 \pm 0.1$ & $0.01 \pm 0.01$ & $0.57 \pm 0.09$\\
Significance &  & $1.8 \pm 0.1$ & $1.8 \pm 0.2$ & 0 & $14 \pm 9$ & 0 & $10 \pm 10$ & $2.6 \pm 0.4$\\ \midrule \midrule 
Classifier LLF AUC & 0.50 & 0.52 & 0.54 & 0.50 & 0.97 & 0.81 & 0.93 & 0.99\\
Classifier HLF AUC & 0.50 & 0.53 & 0.55 & 0.50 & 0.84 & 0.64 & 0.74 & 0.92

%% file: tables/3_model_measurements.tex
 & \multirow{2}{*}{FPD $\times 10^3$} & \multirow{2}{*}{KPD $\times 10^3$} & \multirow{2}{*}{$W^M_1$ $\times 10^3$} & \multirow{2}{*}{\wassppt $\times 10^3$} & Inference time \\ 
 & & & & & ($\mu$s) per jet \\ \midrule
Truth & $0.08 \pm 0.03$ & $-0.006 \pm 0.005$ & $0.28 \pm 0.05$ & $0.44 \pm 0.09$ & \\
MPGAN & $\mathbf{0.30 \pm 0.06}$ & $\mathbf{-0.001 \pm 0.004}$ & $\mathbf{0.54 \pm 0.06}$ & $0.6 \pm 0.2$ & 41\\
GAPT & $0.66 \pm 0.09$ & $0.001 \pm 0.005$ & $0.56 \pm 0.08$ & $\mathbf{0.51 \pm 0.09}$ & 9